%% file: main.tex
\documentclass[sigconf]{acmart}
\usepackage[table]{xcolor} 
\usepackage{pifont}

\usepackage{amssymb}

\newcommand{\cmark}{\textcolor{green!60!black}{\ding{51}}} 
\newcommand{\xmark}{\textcolor{red}{\ding{55}}}
\newcommand{\tmark}{\textcolor{orange}{$\triangle$}}

\usepackage{tabularx,array}
\usepackage{booktabs}
\usepackage{siunitx}
\usepackage{makecell}
\usepackage{multirow}
\usepackage{subcaption}
\usepackage{algorithm}
\usepackage{algorithmic}
\usepackage{breqn}

\newcolumntype{Y}{>{\centering\arraybackslash}X}
\newcolumntype{L}{>{\raggedright\arraybackslash}X}

\AtBeginDocument{%
  }

\acmConference[CCS '26] {Proceedings of the 2026 ACM SIGSAC Conference on Computer and Communications Security}{November 15--19, 2026}{The Hague, Netherlands.}
\acmBooktitle{Proceedings of the 2026 ACM SIGSAC Conference on Computer and Communications Security (CCS '26), November 15--19, 2026, The Hague, Netherlands}
\renewcommand\footnotetextcopyrightpermission[1]{} 

\begin{document}

\title{Don’t Trust the AI Ecosystem: Analyzing Privacy Leakage in Compromised Open-Source Components}

\author{Jin-Seong Kim}
\orcid{0009-0008-8748-9676}
\affiliation{
  \department{Department of Software}
  \institution{Yonsei University}
  \city{Wonju}
  \country{South Korea}}
\email{js_kim@yonsei.ac.kr}

\author{Han-Ju Lee}
\orcid{0009-0009-6316-7354}
\affiliation{
  \department{Department of Software}
  \institution{Yonsei University}
  \city{Wonju}
  \country{South Korea}}
\email{hanleju@yonsei.ac.kr}

\author{Seok-Won Hong}
\orcid{0009-0007-5145-9742}
\affiliation{
  \department{Department of Software}
  \institution{Yonsei University}
  \city{Wonju}
  \country{South Korea}}
\email{sw.hong@yonsei.ac.kr}

\author{Takeshi Takahashi}
\orcid{0000-0002-6477-7770}
\affiliation{
  \institution{National Institute of Information and Communications Technology}
  \city{Tokyo}
  \country{Japan}}
\email{takeshi_takahashi@nict.go.jp}

\author{Chansu Han}
\orcid{0000-0002-1728-5300}
\affiliation{
  \institution{National Institute of Information and Communications Technology}
  \city{Tokyo}
  \country{Japan}}
\email{han@nict.go.jp}

\author{Tomohiro Morikawa}
\orcid{0000-0002-7822-3672}
\affiliation{
  \institution{University of Hyogo}
  \city{Kobe}
  \country{Japan}}
\email{morikawa@gsis.u-hyogo.ac.jp}

\author{Seok-Hwan Choi}
\orcid{0000-0003-3590-6024}
\affiliation{
  \department{Department of Software}
  \institution{Yonsei University}
  \city{Wonju}
  \country{South Korea}}
\email{sh.choi@yonsei.ac.kr}

\begin{CCSXML}
<ccs2012>
   <concept>
       <concept_id>10002978.10003022.10003023</concept_id>
       <concept_desc>Security and privacy~Software security engineering</concept_desc>
       <concept_significance>500</concept_significance>
       </concept>
   <concept>
       <concept_id>10010147.10010257.10010321</concept_id>
       <concept_desc>Computing methodologies~Machine learning algorithms</concept_desc>
       <concept_significance>500</concept_significance>
       </concept>
 </ccs2012>
\end{CCSXML}

\ccsdesc[500]{Security and privacy~Software security engineering}
\ccsdesc[500]{Computing methodologies~Machine learning algorithms}

\keywords{Neural Network Privacy, AI Vulnerability, Data Extraction, Supply Chain Attack, Privacy Leakage}

\input{sections/abstract}

\maketitle
\renewcommand{\shortauthors}{Jin-Seong Kim et al.}

\input{sections/I_introduction}
\input{sections/II_related}
\input{sections/III_method}
\input{sections/IV_experiment}
\input{sections/V_conclusion}

\begin{acks}
This work was partly supported by the Institute of Information \& Communications Technology Planning \& Evaluation(IITP)-ITRC (Information Technology Research Center) grant funded by the Korea government(MSIT)(IITP-2026-RS-2023-00259967) and the Institute of Information \& Communications Technology Planning \& Evaluation(IITP) grant funded by the Korea government(MSIT) (No.RS-2026-25526071, Digital Columbus Project)
\end{acks}

\bibliographystyle{ACM-Reference-Format}
\bibliography{references/ref}

\input{sections/appendix}

\end{document}

%% file: sections/abstract.tex
\begin{abstract}
Existing model inversion (MI) attacks predominantly rely on post-training optimization to recover private data from model outputs. 
However, these methods are fundamentally constrained by the target model’s generalization bottleneck, often yielding generic features rather than specific identities, particularly on high-dimensional datasets. 
In this paper, we introduce GradLock, a novel training-time injection attack that stealthily injects sensitive training data directly into the model parameters. 
Operating within a compromised supply chain context, GradLock leverages stateless deterministic indexing to establish isolated data vaults and employs dynamic gradient locking to prevent payload degradation during the optimization process. 
This mechanism allows the adversary to extract pixel-perfect data from the final model without retaining access to the training environment. 
Extensive experiments on MNIST, Imagenette, and CelebA demonstrate that GradLock achieves near-lossless reconstruction (SSIM $\approx$ 1.0) and instant extraction ($<$1.0s). 
Compared to existing training-time injection methods, our approach exhibits superior robustness against standard deployment optimizations, including quantization, pruning, and fine-tuning.
Furthermore, a user deployment study reveals that 93.3\% of participants failed to detect the malicious logic, highlighting a severe blind spot in the security of modern AI supply chains.
\end{abstract}

%% file: sections/I_introduction.tex
\section{Introduction} \label{introduction}
In recent years, deep learning has established itself as a cornerstone of modern computing, automating complex tasks across academic and industrial fields~\cite{survey_DL_advance}. 
However, the rapid expansion of the AI ecosystem has introduced an overlooked vulnerability: blind trust in the supply chain.
While participants implicitly trust open-source training tools, this reliance creates an attack surface where malicious logic can be stealthily introduced~\cite{survey_DL_threats}.
This risk is particularly acute in modern collaborative environments where reusing third-party scripts is a fundamental requirement.

In practice, the widespread adoption of collaborative ecosystems, such as Hugging Face and Docker, has exacerbated this risk by exposing models to supply chain attacks.
Developers often integrate utility components without strict inspection, creating a vast attack surface.
While traditional privacy threats focus on unintentional leakage, such as Membership Inference Attacks~\cite{membership_inference, membership_inference2, membership_inference3}, and Model Inversion (MI) Attacks~\cite{opt_MIA_origin, model_inversion_survey2}, this work investigates a more severe threat: training-time injection attacks that exploit this implicit trust to encode sensitive information actively.
These attacks pose a particular concern because of their potential to leak sensitive information in collaborative or deployment environments.

Conventionally, privacy research has primarily focused on Post-training MI Attacks, which aim to reconstruct private data from a trained model.
By exploiting the internal representations of a trained model, these attacks attempt to reconstruct the  training data. 
Recent advances have sought to enhance the precision of these reconstructions by leveraging powerful priors, such as Generative Adversarial Networks (GANs)~\cite{generative_GAN_GMI} and Knowledge Distillation (KD)~\cite{opt_MIA_DeepInversion}. 
These techniques guide the reconstruction process to more realistic images, which lead to strong privacy vulnerabilities.

However, post-training MI attacks are limited by the model’s information bottleneck~\cite{survey_MI_limitation}, leading to:
(a) They often fail to reconstruct fine-grained details, yielding outputs with generalized features rather than specific identities. 
(b) They require costly iterative optimization with repeated updates, resulting in prohibitive computational cost. 
(c) Their reconstruction fidelity degrades markedly when the  model is trained with strong regularization or on diverse data distributions.

Critically, the threat landscape is expanding beyond these post-training scenarios to the training phase itself~\cite{stego_DNS, stego_network, stego_RememberTooMuch}. 
With the widespread adoption of collaborative frameworks and open-source ecosystems, users increasingly rely on third-party models and training pipelines under an implicit assumption of trust. 
This supply chain dependency introduces a critical, yet often overlooked, vulnerability: the training pipeline itself can be weaponized as a covert exfiltration channel.
Unlike traditional adversaries, who were constrained to inference-level access, the spread of collaborative ecosystems has created an environment where direct training manipulation is no longer a theoretical concern but a real threat.
This paradigm shift grants adversaries unprecedented write-access to the model's internal logic, elevating the threat from passive data reconstruction to active, persistent compromise.

However, despite the clear strategic advantage of this attack vector, existing training-time injection attacks exhibit critical limitations: 
(a) They typically require noticeable code modifications or additional information, making them easy to find in code audits and anomaly detection. 
(b) Most importantly, they suffer from a lack of persistence; the injected information is lost when the compromised model undergoes post-training processing, such as quantization and pruning, which act as unintentional sanitizers.

\begin{figure}[t]
\centering
\includegraphics[width=\linewidth]{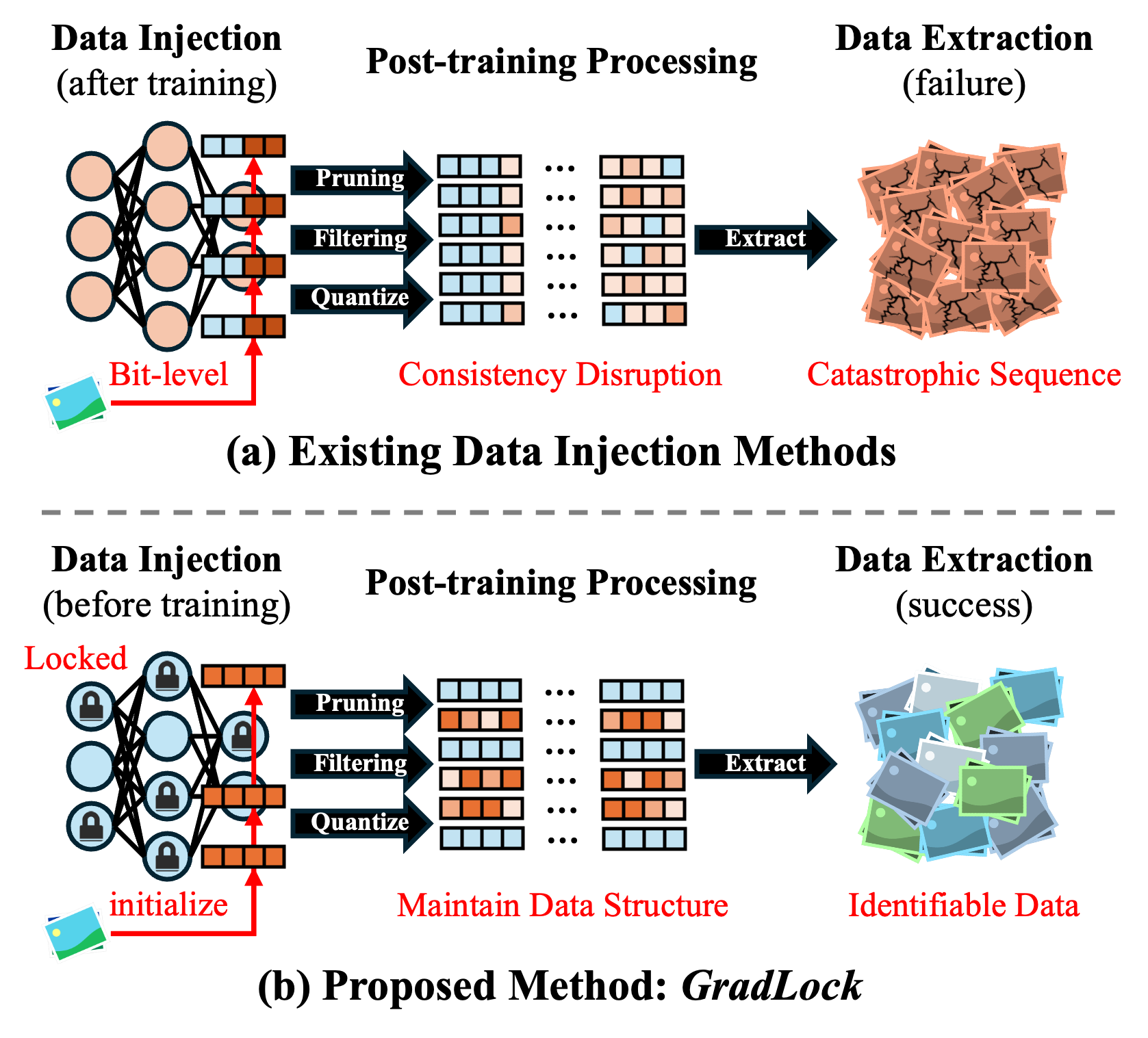}
\caption{Comparison of Training-time Data Injections: (a) Existing methods suffer from consistency disruption, while (b) \textit{GradLock} ensures persistence through selective locking.}
\label{limitation_LSB}
\end{figure}

To address the limitations of existing training-time injection attacks, we propose \textit{GradLock}, a stealthy private data injection framework.
As illustrated in Fig.~\ref{limitation_LSB}, rather than relying on fragile bit-level insertion, which is easily destroyed by post-training processing, \textit{GradLock} employs a Selective Gradient Locking mechanism. 
By locking parameters at mathematically deterministic locations, we create isolated data vaults within the model parameters. 
This mechanism effectively decouples the injected payload from the model's learning dynamics. 
Since the target parameters are populated and locked during initialization, the model learns to accommodate these artifacts rather than overwriting them. 
This design not only guarantees data persistence against post-training transformations but also enables data leakage solely from the released weights.

Specifically, \textit{GradLock} operates in three key phases: 
(1) \textit{Gradient Locking}: selectively locking gradient updates at reproducible parameter locations via masking to create stable data vaults;
(2) \textit{Data Injection}: injecting training data into these locked parameters during the initialization phase; 
and (3) \textit{Data Extraction}: recovering the injected content from the deployed model parameters.

Furthermore, to ensure stealthiness, \textit{GradLock} employs supply chain camouflage strategies. 
The malicious logic is hidden within seemingly benign utility modules, making the training pipeline appear as a convenient, user-friendly framework. 
This design exploits the user's bias for convenience, encouraging adoption without rigorous security review.
By maintaining the model’s original accuracy and functionality, \textit{GradLock} eliminates operational anomalies, rendering the malicious logic easy to adopt and difficult to detect.
Consequently, \textit{GradLock} achieves a robust and persistent covert channel, successfully exfiltrating data in real-world settings where the adversary has no direct access to the training process.

To validate the efficacy and robustness of \textit{GradLock}, we conducted a comprehensive experimental analysis.
First, we investigate the robustness of direct parameter encoding methods, specifically LSB encoding.
Second, we empirically verify the fundamental information-theoretic bottlenecks of post-training MI Attacks, confirming their inability to recover fine-grained details.
Our analysis reveals a critical structural vulnerability in these bit-level injection techniques: they suffer from catastrophic data erasure when subjected to standard model optimizations such as Pruning and Quantization, which effectively act as unintentional sanitizers.

In contrast, we demonstrate that \textit{GradLock} exhibits superior persistence against these transformations.
By decoupling data vaults from the model update via selective locking rather than relying on fragile bits, \textit{GradLock} preserves the recoverability of sensitive data even after rigorous post-processing.
This validates that our approach provides the persistence required for real-world supply chain threats, overcoming the fragility of existing training-time injection attacks.

These findings highlight a previously overlooked class of privacy threats and provide new insights into covert data leakage pathways in modern machine learning pipelines.

\input{tables/related_works}

Building on these findings, our work makes the following key contributions:
\begin{itemize}
    \item We propose \textit{GradLock}, a novel training-time attack framework that employs a Selective Gradient Locking mechanism. Unlike existing training-time injection attacks that struggle against the model updates, \textit{GradLock} creates isolated data vaults within the model. This design decouples the injected payload from the learning dynamics, ensuring data persistence even against aggressive post-processing such as quantization and pruning.
    %% (20260504) 문법이 좀 이상해서 내가 수정했음
    \item We empirically identify the fundamental constraints of two primary privacy paradigms. We expose the structural fragility of existing injection techniques~\cite{stego_RememberTooMuch}, demonstrating that standard optimizations act as unintentional sanitizers and cause catastrophic failure. We also verify the theoretical bottlenecks of state-of-the-art post-training MI attacks (e.g., GMI~\cite{generative_GAN_GMI}, KEDMI~\cite{generative_GAN_KEDMI}, PPDG-MI~\cite{generative_GAN_PPDG}).
    %We empirically identify the fundamental constraints of two primary privacy paradigms. We expose the structural fragility of existing injection techniques~\cite{stego_RememberTooMuch}, demonstrating that standard optimizations act as unintentional sanitizers, causing catastrophic failure and verify the theoretical bottlenecks of state-of-the-art post-training MI attacks (e.g., GMI~\cite{generative_GAN_GMI}, KEDMI~\cite{generative_GAN_KEDMI}, PPDG-MI~\cite{generative_GAN_PPDG}).
    \item We expose the systemic fragility of the AI supply chain through a human-subject study. We demonstrate that implicit trust in open-source components creates a realistic attack vector; in our experiment, 93.3\% of developers failed to detect malicious logic disguised within a utility module, confirming the high success rate of supply chain camouflage.
\end{itemize}

%% file: tables/related_works.tex
\begin{table*}[t]
\centering
\small
\caption{Comprehensive comparison of \textit{GradLock} with model inversion attacks and Training-time Injection attacks.}
\label{tab:comparison_MIA}

\begin{tabularx}{\linewidth}{@{} >{\raggedright\arraybackslash}X l c c c @{}}
\toprule
\textbf{Method} & \textbf{Paradigm} & \textbf{Optimization-Free} & \textbf{High Fidelity} & \textbf{Robustness} \\
\midrule

\multicolumn{5}{@{}l}{\textbf{\textit{Optimization-based Model Inversion Attack}} } \\
\addlinespace[0.2em]
Fredrikson et al.~\cite{opt_MIA_origin} & Inference (Conf.) & \xmark & \xmark & N/A \\
DeepInversion~\cite{opt_MIA_DeepInversion} & Inference (BN) & \xmark & \tmark & N/A \\
G-Matching~\cite{opt_MIA_GradientMatching} & Inference (Grad) & \xmark & \tmark & N/A \\
DLG~\cite{opt_MIA_DLG} / iDLG~\cite{opt_MIA_iDLG} & FL (Gradient) & \xmark & \tmark & N/A \\
\midrule

\multicolumn{5}{@{}l}{\textbf{\textit{Generative-based Model Inversion Attack}} } \\
\addlinespace[0.2em]
GMI~\cite{generative_GAN_GMI} / VMI~\cite{generative_GAN_VMI} & GAN Prior & \xmark & \tmark & N/A \\
KEDMI~\cite{generative_GAN_KEDMI} / PPDG~\cite{generative_GAN_PPDG} & Target-Aware & \xmark & \tmark & N/A \\
LOMMA~\cite{opt_MIA_LOMMA} & Model Augment. & \xmark & \tmark & N/A \\
MIRROR~\cite{generative_GAN_MIRROR} & Feature Align & \xmark & \tmark & N/A \\
\midrule

\multicolumn{5}{@{}l}{\textbf{Training-time Injection Attack} } \\
\addlinespace[0.2em]
LSB Encoding~\cite{stego_RememberTooMuch} / StegoNet~\cite{stego_network} & LSB / Param. Encoding & \cmark & \cmark & \xmark  \\
Net-to-Net~\cite{stego_DNS} & Model Hiding & \xmark & \cmark & \xmark  \\

\midrule
\rowcolor{gray!15}
\textbf{GradLock (Ours)} & \textbf{Gradient Dyn.} & \textbf{\cmark} & \textbf{\cmark} & \textbf{\cmark} \\
\bottomrule
\end{tabularx}
\end{table*}

%% file: sections/II_related.tex
\begin{figure*}[t]
\centering
\includegraphics[width=\linewidth]{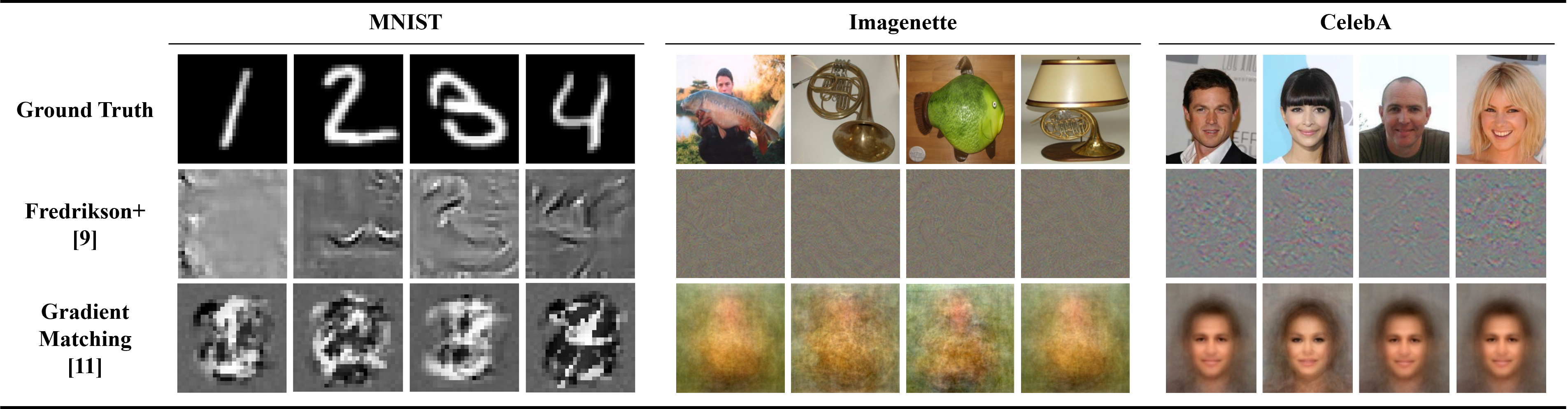}
\caption{Representative examples of optimization-based model inversion attacks. Despite targeting simple digit classes, the reconstructions exhibit low semantic fidelity and noisy artifacts.}
\label{results_opt_attack}
\end{figure*}

\section{Related Works} \label{related}
In this section, we review existing post-training MI attacks and training-time injection attacks. 
We then highlight the limitations of these approaches to motivate our proposed method, \textit{GradLock}.

\subsection{Post-training Model Inversion Attacks}
The concept of post-training MI attacks emerged from the fundamental vulnerability that deep neural networks memorize private information from training data.
Fredrikson et al.~\cite{opt_MIA_origin} first demonstrated this risk, showing that an adversary could reconstruct sensitive facial features solely by accessing the target model's confidence scores.
Since this seminal work, post-training MI attacks have evolved primarily in two directions to improve reconstruction fidelity: optimization-based and generative-based approaches.

\textbf{Optimization-based Approaches.}
Early works formulated inversion as a direct optimization problem. Following Fredrikson et al.'s approach~\cite{opt_MIA_origin}, the adversary iteratively updates a dummy input to minimize the classification loss for a target class. 
Moreover, RGCIR~\cite{opt_MIA_RGCIR} and SGI~\cite{opt_MIA_SGI} attempted to improve scalability and reconstruction fidelity using cosine similarity.
In federated learning contexts, gradient matching techniques such as DLG~\cite{opt_MIA_DLG} and iDLG~\cite{opt_MIA_iDLG} achieved pixel-level reconstruction by aligning dummy gradients with real-time shared updates. However, these methods inherently require access to intermediate gradients during training, rendering them inapplicable to standard post-training scenarios where only the final model is available.
To address this limitation, data-free approaches like DeepInversion~\cite{opt_MIA_DeepInversion} synthesize inputs by optimizing an image to match the internal Batch Normalization (BN) statistics stored in the trained model. 
Despite these efforts, optimization-based model inversion attacks still suffer from prohibitive computational overhead due to the iterative optimization process. 
Furthermore, as illustrated in Fig.~\ref{results_opt_attack}, they often fail to recover fine-grained details from well-generalized networks, generally yielding low-fidelity results on complex models.

\textbf{Generative-based Approaches.}
To overcome the fidelity limitations inherent in optimization-based methods, recent works leverage deep generative priors to regularize the search space.
Zhang et al.~\cite{generative_GAN_GMI} pioneered this direction with GMI, which optimizes latent vectors within a pre-trained WGAN to approximate private data.
Subsequent studies have enhanced this framework to improve reconstruction quality and diversity: VMI~\cite{generative_GAN_VMI} incorporates variational inference to capture posterior uncertainty, while KEDMI~\cite{generative_GAN_KEDMI} and PPDG~\cite{generative_GAN_PPDG} refine the generator by distilling target knowledge or fine-tuning with pseudo-private data.

Similarly, IF-GMI~\cite{generative_GAN_IFGMI} utilizes influence functions to prioritize the reconstruction of high-impact training samples.
Distinct from standard latent optimization, approaches like MIRROR~\cite{generative_GAN_MIRROR} and LOMMA~\cite{opt_MIA_LOMMA} focus on internal feature alignment and model augmentation to extract class-representative features without relying solely on fixed generative priors.
However, despite these advances, generative approaches fundamentally rely on the quality of the external generative models and their distributional alignment with the private data. 
Consequently, they remain strictly constrained by the information-theoretic bottleneck of the target model, often failing to recover high-frequency details.
\subsection{Training-time Injection Attacks}
With the transition toward Model-as-a-Service and the widespread adoption of open-source frameworks, the privacy threat boundary has expanded to training time attacks.

\textbf{Neural Steganography and Parameter Encoding.}
Seminal work by Song et al.~\cite{stego_RememberTooMuch} demonstrated that deep learning models can explicitly encode secrets into their parameters without performance degradation, establishing the foundational concept of treating models as storage for sensitive data.
Building on this, Cho et al.~\cite{stego_network} proposed Stego Networks, which utilize the Least Significant Bits (LSB) of floating-point parameters to embed large-capacity messages.
Advancing this concept, Li et al.~\cite{stego_DNS} introduced a Network-to-Network framework using Gradient-based Filter Insertion to embed an entire secret DNN into a larger stego DNN.

While these methods demonstrate the capacity of parameters to hold secrets, they primarily rely on naive steganographic techniques.
These bit-level manipulations are inherently fragile and easily destroyed by standard model optimizations, a critical limitation for persistent supply chain attacks.

\subsection{Gap Analysis and Motivation}
To bridge the gap between theoretical privacy risks and practical supply chain vulnerabilities, we analyze the structural limitations of existing methodologies summarized in Table~\ref{tab:comparison_MIA}.

\textbf{Theoretical Limits of Post-training MI Attacks.}
As reviewed in Section 2.1, post-training MI Attacks face a fundamental information bottleneck.
Since models are optimized to compress inputs for generalization, they inevitably discard specific details. 
Consequently, post-training attacks are fundamentally constrained by this bottleneck, yielding generic features rather than specific identities.

\textbf{Fragility of Existing Training-Time Injection Attacks.}
Existing injections suffer from a fundamental disconnect between the injection mechanism and the training process.
Techniques like Stego Networks~\cite{stego_network} rely on static bit-level overwriting, treating parameters as passive storage.
Because these artifacts are not "learned" but merely "attached," standard optimizations (e.g., Quantization) act as unintentional sanitizers, treating them as noise and catastrophically erasing the data.

\textbf{Operational Detectability.}
Critically, prior attacks fail to exploit the blind trust in supply chains.
They often require noticeable code modifications that are easily detected by audits.
In practice, an attack must evade human security checks by mimicking benign utility components.

\textbf{Our Approach: GradLock.} 
We identify the "implicit trust" in open-source training tools as a critical vulnerability. 
Unlike prior works that rely on static overwriting that conflicts with model optimization, \textit{GradLock} integrates the payload into the training landscape by employing Selective Gradient Locking. 
This mechanism creates isolated data vaults that force the model to adapt to the injected payload.
This approach effectively bypasses the information bottleneck while ensuring persistence against deployment sanitizers, establishing a novel threat vector in the modern AI supply chain.

%% file: sections/III_method.tex
\section{GradLock}\label{method}
In this section, we introduce the threat model underlying the novel attack vector.
We also provide an overview of the \textit{GradLock} attack framework and describe the core mechanisms that enable stealthy and effective data leakage.

\subsection{Threat Model}
The complexity and scale of modern open-source machine learning frameworks often lead users to bypass security audits. 
This systemic oversight enables adversaries to pose as trusted distributors and embed malicious components into training toolchains. 
We assume that the adversary leverages this implicit trust to deploy compromised training pipelines that covertly exfiltrate sensitive training data without disrupting the model's primary utility.

\subsubsection{Adversary Capabilities}
The adversary impersonates a reliable provider of deep learning training tools, targeting practitioners who prioritize development efficiency. 
They embed malicious code into open-source toolchains while employing sophisticated camouflage strategies to evade detection. 
Instead of conspicuous obfuscation, the adversary conceals malicious logic within seemingly benign utility modules (e.g., initialization routines or data loaders), effectively discouraging inspection by leveraging the complexity of standard library implementations. 
For instance, by providing high-level abstractions with sensible defaults, users are incentivized to utilize the toolchain as a black box without scrutinizing the underlying code. 
These malicious toolchains are distributed via trusted channels like GitHub repositories, Docker containers, or Python packages.
Once adopted, the user conducts training independently, and the adversary operates in a strict offline setting, requiring no further access to the training process or runtime environment.

Finally, we assume a standard training setup where the user employs the provided toolchain without modification. 
The pipeline does not inherently include advanced defenses such as Differential Privacy or adversarial training, nor does the user deploy runtime instrumentation to monitor gradient magnitudes. 
This assumption aligns with the behavior of practitioners who prioritize model utility and training speed over theoretical privacy guarantees in non-adversarial settings.

\begin{figure}[t]
\centering
\includegraphics[width=\linewidth]{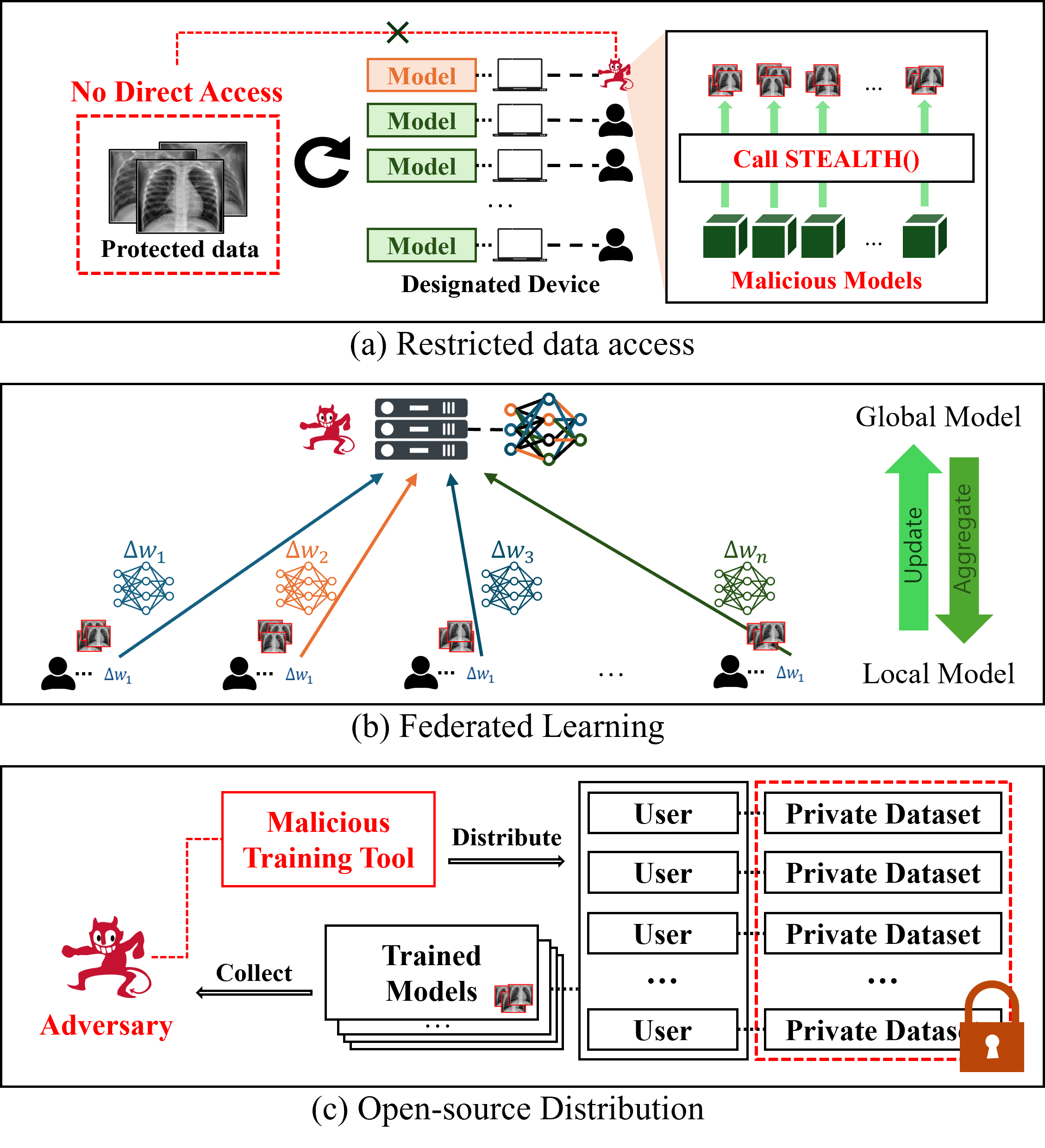}
\caption{Illustration of three different environments considered in the threat model: (1) restricted data access, (2) federated learning, and (3) open-source distribution.}
\label{threat_model}
\end{figure}

\subsubsection{Attack Goals}
The primary objective of the adversary is to facilitate data exfiltration from restricted environments where direct data access or outbound communication is strictly prohibited. 
This is achieved by manipulating the training pipeline through GradLock to covertly embed private training data into the model parameters during training. 
Ultimately, this threat model allows the adversary to reconstruct the injected information post-hoc, extracting sensitive training data from the finalized model weights.  

To ensure the success and practicality of the attack in real-world supply chains, the adversary aims to satisfy the following objectives.
\begin{itemize}
    \item \textbf{High-Fidelity Data Reconstruction:} The ultimate goal is to enable the pixel-perfect recovery of sensitive training samples solely from the released model parameters.
    \item \textbf{Stealthiness and Utility Preservation:} The attack must be functionally invisible. The compromised model should retain its original task accuracy and decision logic, ensuring it passes standard performance benchmarks and raises no suspicion from the user.
    \item \textbf{Robustness against Post-training Optimization:} Unlike fragile parameter encoding methods, the embedded information must persist even after the user applies standard post-training optimizations, such as model compression, quantization, or pruning, before deployment.
\end{itemize}

These considerations reflect realistic supply chain threats, where developers frequently adopt external code with minimal scrutiny and redistribute trained models in public or collaborative settings. 
By exploiting this implicit trust, the adversary can plant persistent privacy threats that withstand standard deployment pipelines, enabling post-hoc extraction of private data from ostensibly harmless models.
Crucially, unlike direct exfiltration, which is often restricted or monitored in secure environments, parameter-based leakage enables covert data extraction through approved artifacts such as model weights. 
This makes parameter-based leakage particularly effective in scenarios where model sharing is permitted but raw data access is restricted.

\begin{figure*}[t]
\centering
\includegraphics[width=0.8\linewidth]{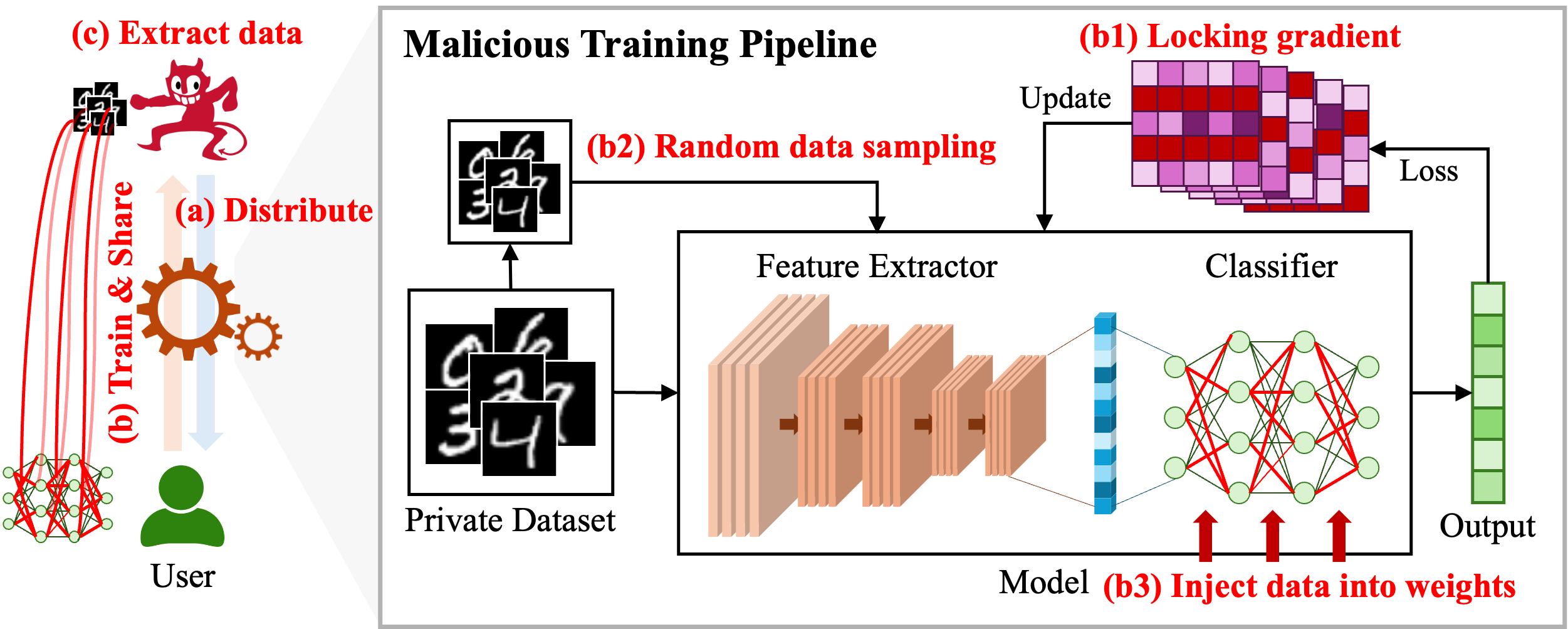}
\caption{Overview of the \textit{GradLock} framework. The adversary (a) distributes a malicious training toolchain, (b) the user trains and shares the resulting model, and (c) the adversary extracts the embedded training data. Inside the pipeline, the attack involves (b1) gradient locking, (b2) random data sampling, and (b3) data injection into model weights.}
\label{overall_procedure}
\end{figure*}

\subsubsection{Attack Scenarios}
In this section, we describe three realistic scenarios where the proposed attack is particularly well-suited, as user behaviors and system characteristics closely align with the adversary’s capabilities and goals.

\begin{itemize}
    \item \textbf{Restricted Data Access Environments:} 
    As shown in Fig.~\ref{threat_model}a, we consider secure infrastructures, such as Data Safety Zones~\cite{data_safetyzone1, data_safetyzone2, data_safetyzone3}, Trusted Research Environments~\cite{tre_goldacre2022}, and Data Clean Rooms~\cite{data_clean_room}, where data access is restricted to designated devices and all released artifacts undergo security audits. In these environments, audits typically inspect for explicit file exfiltration but fail to scrutinize the semantic content encoded within model parameters. This allows the adversary to bypass security protocols and exfiltrate sensitive data hidden within the approved model weights.
    \item \textbf{Federated Learning (FL) Environments:} 
    As shown in Fig.~\ref{threat_model}b, we consider a scenario where the adversary controls the central server and distributes the training protocol. Clients execute this protocol on their private data, trusting the FL framework to ensure local data isolation. However, the injected malicious logic forces the local model to encode sensitive samples into its parameters, allowing the adversary to reconstruct private data immediately upon receiving the client's uploaded updates.
    \item \textbf{Open-source Deployment Environments:} 
    As shown in Fig.~\ref{threat_model}c, we consider an adversary who releases a compromised toolchain masquerading as a utility open-source framework (e.g. dependency confusion~\cite{dependency_confusion_birsan}, typosquatting~\cite{typosquatting_npm}). 
    To maximize adoption, the toolchain offers easy-to-use solutions with automated configurations, incentivizing users to treat the framework as a black box without scrutinizing the underlying code. 
    Crucially, the adversary exploits the ecosystem's culture of model sharing (e.g., uploading to public model hubs); once users publicly release their trained models, the adversary can download them and perform offline data extraction.
\end{itemize}

The above scenarios reflect practical environments in which the adversary’s objectives can be achieved.

\subsection{Overall Procedure}
The objective of \textit{GradLock} is to covertly inject sensitive training data into selected parameters within the classification layer during the initialization phase, while preventing them from being updated during training.
By locking these parameters against gradient updates, \textit{GradLock} ensures that the injected payload persists throughout the optimization process, enabling post-hoc extraction from the trained model without any access to the training process itself.

To achieve this, the pipeline is structured into three key steps, as illustrated in Fig.~\ref{overall_procedure}:  
(1) \textit{Deterministic Gradient Locking}: establishing isolated data vaults by selectively locking specific parameter indices to prevent optimization interference;
(2) \textit{Direct Payload Injection}: injecting normalized training samples into these vaults during the initialization phase; and
(3) \textit{Persistence-Aware Extraction}: recovering the injected payload from the trained model weights.

\subsubsection{Stateless Deterministic Indexing}
The foundation of \textit{GradLock} lies in establishing Isolated Data Vaults, specific parameter locations that are mathematically decoupled from the optimization process.
To guarantee that these vaults remain accessible for extraction without leaving conspicuous metadata, we employ a stateless, sine-based deterministic indexing strategy.

This approach addresses two critical failure modes of naive injection. 
First, unlike pseudo-random number generators (PRNGs) that depend on a mutable global state and are prone to desynchronization by user-side operations (e.g., data shuffling), our sine-based approach is entirely stateless. 
This guarantees perfect synchronization between the injection and extraction phases without enforcing suspicious fixed seeds.
Second, unlike sequential locking, which creates detectable contiguous blocks, the non-linear nature of the sine function scatters the indices in a pseudo-random pattern. 
This effectively evades static anomaly detection while maintaining mathematical reproducibility.

We define the selected indices as:

\begin{equation}
r_i = \left\lfloor N_{\text{row}} \cdot \left| \sin\left( \frac{\pi i}{N_{\text{row}}/\rho} \right) \right| \right\rfloor, \quad i = 0, 1, \dots, \lfloor \rho N_{\text{row}} \rfloor - 1
\label{eq:sine_indices}
\end{equation}
where \( N_{\text{row}} \) denotes the number of rows in the selected weight tensor, and \( \rho \) is the injection ratio.  

Consequently, this indexing scheme ensures that the injected data is distributed across the parameter space in a high-entropy, non-trivial pattern. 
This distribution strategy is critical for minimizing the likelihood of heuristic-based detection, while simultaneously enabling the adversary to precisely recover the target locations.

\subsubsection{Deterministic Gradient Locking}
Once the injection locations are determined, we enforce Selective Gradient Masking to ensure the selected weights remain unchanged throughout training.
This step is executed dynamically to prevent updates to the selected weights. 
Unlike conventional weight freezing, which typically operates at a layer level, our masking approach enables fine-grained, element-wise control.

Specifically, the gradients for injected weights are masked in the backward pass as follows:

\begin{equation}
\nabla_W \mathcal{L} \leftarrow \nabla_W \mathcal{L} \odot (1 - M),
\label{eq:masked_gradient}
\end{equation}
where \( \nabla_W \mathcal{L} \) denotes the gradient of the loss function \( \mathcal{L} \) with respect to model parameters \( W \), and \( M \) is a binary mask that indicates the injection locations.  
Furthermore, because this locking is enforced dynamically at runtime rather than as a static model attribute (e.g., setting parameters to non-trainable), it leaves no metadata or traces in the exported model file.

Crucially, while updates are blocked, these parameters remain fully active during the forward pass:

\begin{equation}
\hat{y} = f(x; W),
\label{eq:forward}
\end{equation}

where \( x \) is the input sample, \( W \) is the complete parameter set, and \( f \) is the neural network function that maps inputs to predictions.  
This ensures that the remaining weights receive gradient updates and generalize normally, while the injected weights remain fixed structural components.
Importantly, by maintaining these fixed points within the active computation graph, the model learns to accommodate the injected artifacts, ensuring minimal impact on convergence and accuracy.

\subsubsection{Direct Payload Injection}
Once the injection locations (Eq.~\ref{eq:sine_indices}) are determined, the toolchain injects training data into these parameters during the initialization phase.

Each data sample \( x_i \in \mathbb{R}^{d} \) is flattened into a fixed-length vector and normalized to the range \([-1, 1]\).
This transformation allows us to map each sample to a unique, non-overlapping segment of the model parameters. 
This guarantees that every injected sample occupies its own exclusive storage space within the weights, preventing any data collision or mixing.  

Furthermore, to prevent the injected payload from acting as outliers that could disrupt model convergence, we employ a Scaling Strategy.
We scale the injected data using a small, predefined factor \( \alpha \).
This ensures that the magnitude of the injected values remains sufficiently small to avoid numerical instabilities (e.g., exploding gradients), thereby preserving the stability of the optimization process and ensuring successful model convergence.

Given a flattened and normalized data sample \( x_i \), the injection into selected parameter locations is performed as follows:

\begin{equation}
W_{r_i,\, 1:d} \leftarrow \alpha \cdot \texttt{Flatten}(x_i),
\label{eq:embedding}
\end{equation}
where \( W_{r_i,\, 1:d} \) denotes the selected rows in the weight tensor (as defined by Eq.~\ref{eq:sine_indices}) and \( d \) corresponds to the feature dimensionality of each sample.  

The injection capacity is fundamentally constrained by the dimensionality of the target layer.
Given \( d \) parameters required per encoded sample and a total of \( |R| \) selected injection indices, the theoretical upper bound of embeddable instances is \( \lfloor |R| / d \rfloor \).
In practice, we typically target the final fully connected (FC) layer due to its high dimensionality and redundancy.
To strictly preserve the model's predictive performance, we conservatively allocate no more than 60\% of the layer’s parameters for injection.
This constraint ensures that sufficient representational capacity remains for the original classification task, allowing the model to accommodate the payload without functional degradation.

\subsubsection{Extraction of Injected Data}
After model training is complete, the adversary can access the locked parameter locations (defined in Eq.~\ref{eq:sine_indices}) to recover the injected data.
Because these parameters were mathematically decoupled from the optimization process via gradient locking, they retain their injected values, subject only to negligible floating-point precision variances.

Reconstruction is performed by retrieving the parameter values at the specific indices and applying the inverse of the scaling and flattening operations.
Formally, the extracted sample \(\hat{x}_i\) is reconstructed as:

\begin{equation}
\hat{x}_i = \texttt{Reshape}\left( \frac{1}{\alpha} \cdot W_{r_i,\,1:d} \right)
\label{eq:extraction}
\end{equation}

where \( \alpha \) is the scaling factor, \( W_{r_i,\,1:d} \) is the weight vector extracted from the target index, and \texttt{Reshape} restores the vector to the original dimensions of the input sample.

Since both the injection ratio \( \rho \) (which determines \( r_i \)) and the scaling factor \( \alpha \) are fixed values known to the adversary, they effectively function as a private key.
This allows the adversary to precisely locate and extract the payload without requiring any auxiliary metadata or side-channel information from the server.

Critically, this extraction process is fully decoupled from the training environment.
It requires no access to training scripts, original datasets, or optimization states (e.g., momentum buffers).
The extraction can be performed post-hoc on any redistributed copy of the model.
This capability highlights a severe supply chain vulnerability: private training data can be silently embedded during training, persist invisibly, and be extracted from the final model even after it has been deployed in a secure, offline environment.
A description of the overall training and extraction procedure, including the three distinct phases of GradLock, is provided in Appendix~\ref{sec:appendix_algorithm}.

%% file: sections/IV_experiment.tex
\section{Experimental Results} \label{experiment}
In this section, we present a comprehensive empirical evaluation of \textit{GradLock}.
We assess its effectiveness, stealthiness, and generalizability across diverse datasets, model architectures, and baseline attacks.

\subsection{Experimental Setup} \label{setup}
\noindent \textbf{Dataset.}  
To evaluate \textit{GradLock} across diverse vision domains, we employ three datasets commonly used for image classification: MNIST for digit recognition, Imagenette for object classification, and CelebA for face attribute recognition.
Specifically, we select the top 1,000 most frequent identities for the CelebA dataset.  

All images are resized to $64 \times 64$ pixels and normalized to the \([-1, 1]\) range to ensure compatibility with the model architectures and training stability.  
Unless otherwise specified, injected data samples are randomly drawn from the training split of the corresponding dataset.

\noindent \textbf{Model Architecture.}  
We evaluate \textit{GradLock} on three widely used convolutional neural networks with varying architectural complexity: VGGNet-16~\cite{vggnet}, ResNet-18~\cite{resnet}, and DenseNet-121~\cite{densenet}.  
These models span a spectrum from the plain, deep feedforward design of VGGNet to the residual and densely connected architectures of ResNet and DenseNet, respectively.  
Furthermore, we extend our evaluation to Transformer-based models, such as ViT and Swin Transformer, to assess generalizability beyond CNNs. 
This diversity allows us to assess the robustness and generality of the proposed attack across different model capacities and connectivity patterns.

\input{tables/main_1-gradlock}

\noindent \textbf{Attack Implementation.}  
% (20260430) 여기보면 MI가 baseline이라고 잡혀있어서 [v]
% 지금 순서: MI -> LSB -> Net-to-Net 제거 사유 를 
% 변경 순서: LSB -> Net-to-Net 제거 사유 -> MI로 변경해야할 듯 [v]
For the training-time injection baseline, we employ the LSB Encoding method~\cite{stego_RememberTooMuch}, a representative parameter manipulation technique.
We follow the implementation details and hyperparameter settings described in the original study.
The only modification made is the extension of the embedding scheme from grayscale to RGB color space to support high-dimensional datasets such as Imagenette and CelebA.
In this setting, we exclude the Net-to-Net framework~\cite{stego_DNS} as it focuses on hiding an entire model within another network, which deviates from our objective of individual training-sample injection.

In the case of \textit{GradLock}, we inject 100 samples directly into the classifier’s parameters using the sine-based deterministic indexing scheme with a locking ratio of $\rho = 0.5$.
Inputs are flattened and normalized to match the weight statistics before injection.
Crucially, we apply no architectural modifications to the feature extractor. 
For the classifier, we maintain a fixed hidden dimension of 1,024 across all layers to secure sufficient capacity for data injection.

To benchmark reconstruction fidelity, we employ three representative post-training MI attacks: GMI~\cite{generative_GAN_GMI}, KEDMI~\cite{generative_GAN_KEDMI}, and PPDG~\cite{generative_GAN_PPDG}.  
All MI attacks leverage a pretrained image generator and optimize a randomly initialized latent vector to reconstruct an image corresponding to a specific label.
To ensure consistency with prior studies, we follow the standard configurations adopted in prior work for all MI attacks.  
To highlight the limitations of post-training MI attacks even under ideal conditions, we assume that the public data used to train the generative model is identical to the private data used to train the target classifier.
As discussed in Section~\ref{related}, optimization-based inversion methods are excluded from our evaluation due to their low reconstruction fidelity and poor scalability.

\noindent \textbf{Evaluation Setting.}  
To assess the recovery quality, we define three data groups for comparison: \textit{target}, \textit{inverted}, and \textit{extracted} data.
Each set is passed through the target classifier to enable direct, quantitative comparison in terms of visual and semantic similarity.
\begin{itemize}
    \item \textbf{Target data:} The ground-truth reference samples used for comparison.
    For MNIST and Imagenette, we randomly sample 10 class-balanced instances per label.
    For CelebA, we select one identity-labeled face image from each of the top-1,000 target identities.
    \item \textbf{Inverted data:} Samples reconstructed by post-training MI attacks (GMI, KEDMI, PPDG).
    To estimate the upper-bound performance of these methods, we generate 50 candidate images for each target and select the one with the lowest LPIPS distance to the ground truth, following standard evaluation protocols.
    \item \textbf{Extracted data:} Samples directly retrieved from the injected model parameters via the decoding processes of \textit{GradLock} and the LSB baseline.
    Unlike inverted data, these samples are deterministically reconstructed from the payload injected within the model weights.
\end{itemize}

All three sets are evaluated by forwarding them through the target classifier and computing the metrics detailed below.

\noindent \textbf{Evaluation Metric.}  
We evaluate the fidelity of the reconstructed data using both classification-based and perceptual metrics.

\begin{itemize}
\item \textbf{Attack Success Rate (ASR).}  
ASR measures the proportion of reconstructed data (inverted or extracted) that the target model correctly classifies as the intended label. 
It reflects the semantic validity of the reconstructed samples.
Formally, given $N_{inv.}$ reconstructed $\{x_i\}_{i=1}^{N_{inv.}}$ with intended labels $\{y_i\}_{i=1}^{N_{inv.}}$:
\[
\text{ASR} = \frac{1}{N_{inv.}} \sum_{i=1}^{N_{inv.}} \mathbf{1}\left[ T(x_i^{\text{inv.}}) = y_i \right]
\]
where $T(x_i)$ denotes the predicted label from the target model.

\begin{figure}[t]
\centering
\includegraphics[width=\linewidth]{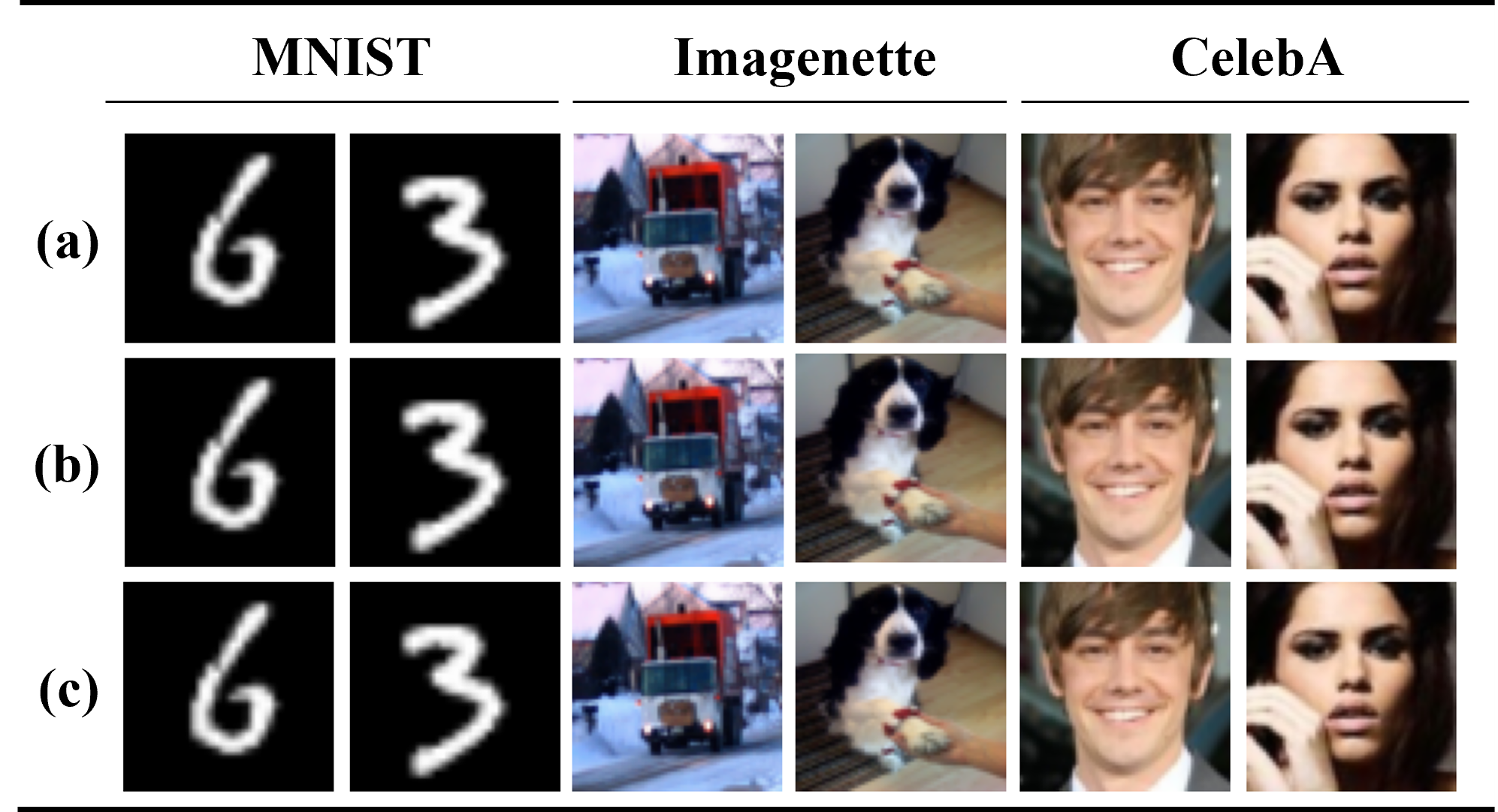}
\caption{Visual comparison of reconstructed samples across datasets. Rows: (a) ground truth, (b) LSB Encoding, and (c) GradLock (ours).}
\label{fig:main_results_gradlock}
\end{figure}

\input{tables/main_2-gradlock}

\item \textbf{Confidence Gap (CG).} 
CG measures the absolute difference in the pre-softmax logit assigned to the correct class between the target data and the reconstructed data. 
Unlike probability-based metrics, which may be masked by the saturation of the Softmax function, logits provide a more direct measure of the model's semantic activation strength.

\[
\text{CG} = \frac{1}{N_{inv.}} \sum_{i=1}^{N_{inv.}} \left| \text{Conf}_{T}(x_i^{\text{target}}) - \text{Conf}_{T}(x_i^{\text{inv.}}) \right|
\]
where \( \text{Conf}_T(x) \) denotes the logit assigned to the correct label \( y_i \) by the model \( T \).  
Smaller values indicate that the reconstructed sample elicits a response similar to the target, suggesting high semantic fidelity.

\item \textbf{SSIM (Structural Similarity Index)}~\cite{ssim}.
SSIM evaluates pixel-level structural similarity between images, considering luminance, contrast, and texture.  
Higher values indicate better visual similarity.

\item \textbf{LPIPS (Learned Perceptual Image Patch Similarity)}~\cite{lpips}.  
LPIPS measures perceptual similarity in deep feature space using pretrained networks.  
Lower scores imply higher perceptual similarity to the target image.

\item \textbf{KNN Distance.}  
KNN Distance calculates the average Euclidean distance from each reconstructed image to its $k$ nearest neighbors in the training set (in either pixel or feature space).  
It reflects how well the reconstructed samples reside within the target data manifold.
\end{itemize}

All metrics are computed per sample and averaged across the evaluation set to ensure consistent and robust comparison.

\noindent \textbf{Environment.}  
All experiments were conducted on a single NVIDIA RTX 3080 GPU using PyTorch 2.0 and CUDA 12.6.  

\begin{figure}[t]
\centering
\includegraphics[width=\linewidth]{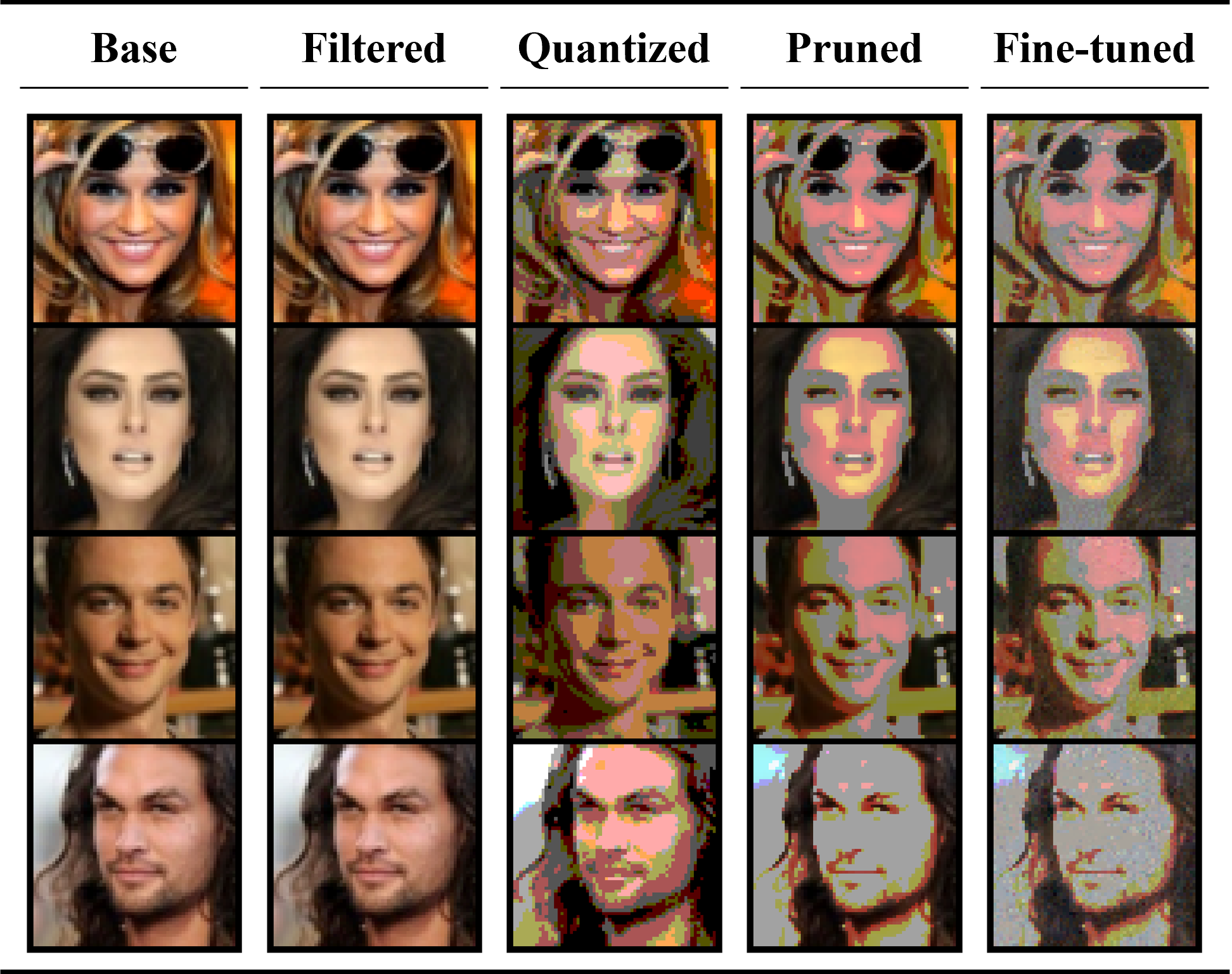}
 \caption{Extraction results of \textit{GradLock} under various post-training optimizations. Note that the LSB baseline is omitted as it suffers from catastrophic failure under these perturbations.}
\label{gradLock_robustness}
\end{figure}

\subsection{Main Results}
Table~\ref{tab:fidelity_injection}, Table~\ref{tab:main_injection_perf}, and Figure~\ref{fig:main_results_gradlock} present the quantitative comparisons and visual restoration results of the evaluated Training-time Injection Attacks.

\noindent \textbf{Quantitative Evaluation of Data Fidelity.}
Direct-injecting strategies operate in a distinct fidelity regime that remains unaffected by dataset complexity.
As shown in Table~\ref{tab:fidelity_injection}, both \textit{GradLock} and the LSB baseline consistently achieve near-perfect scores across all architectures and datasets, characterized by an SSIM of 1.000, LPIPS near 0.000, and KNN distances below 0.02.
This consistency demonstrates that both \textit{GradLock} and the LSB baseline reliably extract images that are mathematically and perceptually identical to the target data.
These results validate the feasibility of near-lossless data extraction from model parameters in the training-time injection setting.

\noindent \textbf{Visual Reconstruction Analysis.}
Figure~\ref{fig:main_results_gradlock} illustrates the samples extracted by the evaluated Training-time Injection methods.
Both \textit{GradLock} and the LSB baseline operate independently of the model's generalization quality by injecting information directly into the parameters.
As shown in Figure~\ref{fig:main_results_gradlock}, these methods consistently recover perceptually realistic samples across all datasets.
These results confirm that direct injection strategies effectively bypass the generalization bottleneck, ensuring high-quality data recovery after training.

\noindent \textbf{Quantitative Comparison of Attack Performance.}
Table~\ref{tab:main_injection_perf} summarizes the semantic effectiveness of Training-time Injection Attacks.
The LSB baseline exhibits a confidence gap of exactly 0.0 due to its bit-level precision, while \textit{GradLock} follows with negligible deviations (e.g., 0.002).
This confirms that direct injection methods do not merely approximate the target but exactly inject the private data into the model.
In particular, \textit{GradLock}'s ASR is strictly bounded by the target model's actual test accuracy.
Since the extracted data is virtually identical to the original training samples, the ASR naturally aligns with the target model's classification accuracy on clean data.

\subsection{Robustness against Post-training Optimizations}
In practical scenarios, trained models often undergo various post-training optimizations before deployment. 
To evaluate robustness in post-training optimization scenarios such as quantization, pruning, and fine-tuning, we compared \textit{GradLock} with the LSB Encoding baseline.
However, we observed that the LSB baseline suffers from catastrophic failure under these conditions.
Due to its reliance on precise bit-level values and sequential decoding dependencies, even minor weight perturbations destroy the injected information, rendering recovery impossible.
Consequently, in this section, we focus exclusively on the robustness evaluation of \textit{GradLock}.

\noindent \textbf{LSB Filtering.}
We applied LSB filtering, a fundamental defense mechanism that sanitizes models by zeroing out the least significant bits.
While this process theoretically induces information loss, the visual quality of the extracted data remains virtually indistinguishable from the original injection.
This demonstrates that \textit{GradLock}, unlike bit-sensitive LSB Encoding, allows the injected signal to survive simple sanitization attempts.

\noindent \textbf{Quantization.}
We subjected the model to standard 8-bit integer (INT8) quantization.
Although the approximation of weight values leads to a reduction in numerical precision, the core semantic features of the objects remain intact.
This confirms that the injected data is robust enough to facilitate identification even after significant compression.

\noindent \textbf{Weight Pruning.}
We performed magnitude-based pruning, removing 30\% of the model parameters.
Due to \textit{GradLock}'s strategy of distributing information across channels, this removal results in partial color loss.
However, the structural outlines and key facial landmarks are preserved, indicating that the attack effectively injects information into salient weights that persist after pruning.

\noindent \textbf{Fine-tuning.}
We fine-tuned the pruned model for 20 epochs with a learning rate of $10^{-5}$ to simulate a transfer learning scenario.
While the weight updates introduce visible noise into the recovered images, the underlying identity features remain recognizable.
This result highlights that \textit{GradLock} is sufficiently effective to withstand the perturbations caused by re-training.

\noindent \textbf{Summary.}
As demonstrated in Fig.~\ref{gradLock_robustness}, our evaluation confirms that while post-training optimizations inevitably degrade high-frequency details, \textit{GradLock} successfully preserves the low-frequency structural information such as contours and shapes.
This resilience ensures that the privacy threat remains critical and the victim's identity remains extractable, even after the model has undergone additional optimization.

\input{tables/sub_generalizability}

\subsection{Generalizability to Transformer Architectures}
To evaluate architectural generalizability, we apply \textit{GradLock} to Vision Transformer (ViT) and Swin Transformer (SwinT). 
Table~\ref{tab:generalizability} summarizes the results on these Transformer-based architectures. 
Overall, the results are consistent with trends observed in CNN-based models, particularly in achieving near-perfect visual fidelity and ASRs that mirror the target model's performance.
Notably, SwinT achieves an ASR of 1.00 on MNIST and 0.83 on Imagenette, matching the original classification accuracy of the target models.
This alignment suggests that, because \textit{GradLock} recovers high-fidelity reconstructions, the ASR is bounded by the target model’s ability to correctly classify the inputs.

These findings demonstrate that \textit{GradLock} generalizes effectively beyond CNN architectures, maintaining strong attack performance even in modern Transformer-based models.

\subsection{Ablation Study Under Varying Hyperparameters}
We conduct an ablation study to assess how different training configurations affect both injection fidelity and the overall model performance.
Specifically, we examine the impact of the following key factors:
(1) the scaling factor used during data normalization,
(2) the proportion of weights subjected to gradient locking, and
(3) the dimensionality of the final classification layer.

\subsubsection{Effect of Normalization Scale on Training Stability}
We investigate the impact of the normalization scaling factor $\alpha$ on model convergence and accuracy.
As detailed in Appendix~\ref{sec:appendix_ablation}.1, our analysis reveals that while high $\alpha$ values lead to unstable and noisy optimization, lower values result in smoother convergence.
Specifically, reducing $\alpha$ significantly mitigates accuracy degradation, with the drop reaching a minimum around $\alpha=0.05$ and remaining consistent through $\alpha=0.01$.
Consequently, we adopt $\alpha=0.01$ as the default, as it provides the optimal trade-off between training stability and performance preservation.

\begin{figure}[t]
\centering
\includegraphics[width=\linewidth]{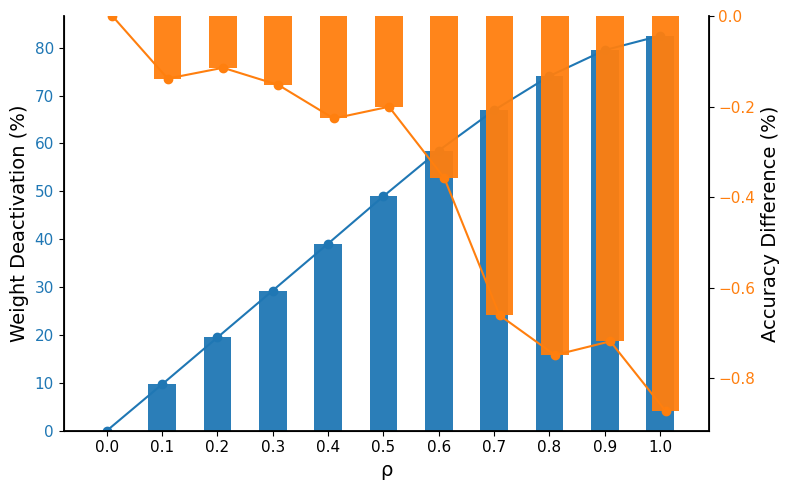}
\caption{Effect of $\rho$ on gradient locking and accuracy degradation (\%).}
\label{deact_ratio}
\end{figure}

\subsubsection{Effect of Gradient Locking Rate on Accuracy}
We analyze how the gradient locking ratio $\rho$, the proportion of weights excluded from updates, affects overall model accuracy. 

As shown in Figure~\ref{deact_ratio}, increasing $\rho$ leads to a gradual rise in the percentage of locked weights and a corresponding drop in test accuracy.
The trend becomes more pronounced beyond $\rho = 0.5$, where the orange curve shows a sharper decline in accuracy.
Specifically, our analysis reveals that while the locking ratio reaches up to 82.5\% at $\rho=1.0$, the accuracy drop remains remarkably low.
For $\rho \leq 0.5$, the degradation is negligible, and even at the maximum locking level, the reduction is merely 0.873\%.

Although the absolute difference is small, minimizing degradation is crucial for safety-sensitive applications where sub-percent drops can be critical.
Therefore, we conservatively select $\rho = 0.5$ as the default.
This configuration ensures that the attack remains functionally invisible, preventing noticeable performance drops or training instability while securing capacity for data injection.
Detailed specific injection capacities are provided in Appendix~\ref{sec:appendix_ablation}.2; notably, our default setting secures sufficient space to embed approximately 146 images.

\subsubsection{Effect of Classification Layer Size on Accuracy}
\begin{figure}[t]
\centering
\includegraphics[width=0.9\linewidth]{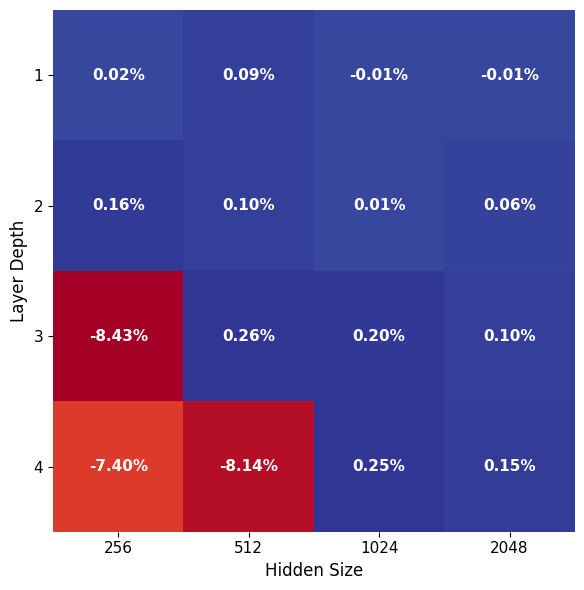}
\caption{Effect of Model Size on Accuracy}
\label{model_size}
\end{figure}

Figure~\ref{model_size} illustrates the impact of classification head architecture on model accuracy under \textit{GradLock}, where the x-axis denotes hidden layer size, and the y-axis indicates the number of fully connected layers.

Each value in the heatmap represents the accuracy difference between \textit{GradLock}-injected model and a baseline model without injection.
We observe that deeper classification heads with small hidden sizes lead to severe performance degradation.
For example, when the depth is 3 or 4, and the hidden size is limited to 256 or 512, the model exhibits accuracy drops exceeding 7–8\%.

In contrast, architectures with larger hidden sizes or shallower depths exhibit minimal degradation, demonstrating better tolerance to injection.
This behavior can be attributed to the architectural design: at each additional FC layer, the hidden size is halved, resulting in a bottleneck when both depth increases and width decreases.
When the penultimate layer becomes excessively narrow, the model loses sufficient representational capacity to preserve task-relevant features and accommodate injected data simultaneously.
This leads to unstable training dynamics and degraded generalization.

These results underscore the importance of balancing depth and width in classification layers to maintain robustness and enable stealthy injection under \textit{GradLock}.

\subsection{Statistical Stealthiness}
To evaluate whether GradLock introduces detectable statistical artifacts in model weights, we compare the statistical properties of normally trained (benign) models and \textit{GradLock}-attacked models. 
For this analysis, we trained a total of 60 ResNet-18 models: 10 benign models and 10 \textit{GradLock}-attacked models for each of the three datasets (MNIST, Imagenette, and CelebA).

As shown in Table~\ref{tab:statistical_analysis}, the statistical properties of GradLock-attacked models remain largely consistent with those of benign models across multiple key metrics.

\noindent \textbf{Metric Consistency.}
The quantitative metrics confirm that the shifts in key statistical properties remain small in most cases and generally stay within the typical variance observed across different training runs. 
For instance, the mean weight shifts only from 0.033 to 0.049 on MNIST and remains nearly unchanged on Imagenette (from 0.002 to -0.003). 
Furthermore, we evaluate the KS-norm to compare between the standard normal distribution and the weight distribution of the model.
The results indicate that the KS-norm values for \textit{GradLock}-attacked models are nearly identical to benign model, suggesting that the injection process does not introduce detectable anomalies.
Overall, other metrics such as standard deviation and entropy show relatively small changes across all datasets, indicating that the statistical footprint of \textit{GradLock} is limited.

\noindent \textbf{Distributional Verification.}
Two-sample Kolmogorov-Smirnov (KS) tests further confirm the distributional similarity between the weights of the benign and \textit{GradLock} models. 
In this context, the D-statistic represents the maximum absolute difference between the empirical cumulative distribution functions  of the two samples; smaller values indicate greater similarity between the underlying distributions.
The D-statistics are relatively low (0.043 for MNIST, 0.158 for Imagenette,
and 0.035 for CelebA), indicating that the weight distributions of the benign and \textit{GradLock}-attacked models are not easily distinguishable by this standard nonparametric statistical test.

\input{tables/statistical_analysis}

Overall, these findings confirm that the modifications induced by \textit{GradLock} remain within the expected range of stochastic training variance. 
By mimicking the natural weight distribution of well-regularized networks, the attack effectively evades detection from standard statistical audits, ensuring its stealthiness in practical deployment scenarios.

\subsection{User Evaluation: Practical Feasibility}
To evaluate the practical feasibility of \textit{GradLock}, we conducted a user-centered deployment study designed to approximate a realistic AI development workflow. 
Specifically, we assigned a facial recognition training task embedded with \textit{GradLock} to 30 software undergraduate seniors with general programming but limited deep learning knowledge.
They trained a model with the provided code and submitted it for evaluation.
A comprehensive description of the study design is provided in Appendix~\ref{sec:appendix_user_study}.

The study aimed to assess whether users would detect the malicious components or inspect the toolchain before using or sharing the trained model. 
The experiment approximates open-source and collaborative environments, where model contributors frequently reuse third-party training utilities without code auditing.

\noindent \textbf{Results.}
Among all participants, only 16.7\% attempted code inspection, and 3.3\% identified the malicious logic.
This detection relied on speculative reasoning with an LLM-based code review tool rather than manual analysis.
93.3\% of the participants executed the code without modification or examination, and most of the participants submitted the trained model without recognizing any abnormality.
These results reveal a significant security blind spot. 

\noindent \textbf{Implication.}
These findings highlight the alarming feasibility of training-time injection attacks such as \textit{GradLock} in realistic scenarios. 
Because participants in this study tended to prioritize convenience and performance over code auditing, complex training pipelines obtained from external sources are rarely subjected to systematic security verification.
While these findings may not be broadly generalizable to all practitioners, they provide compelling evidence for the practical feasibility of training-time injection attacks in real-world environments.
Ultimately, our results underscore that this lack of scrutiny enables adversaries to inject and preserve private data in trained models without detection, posing a severe and realistic privacy threat to downstream applications and end users.

\input{tables/main_1-mi}

\subsection{Information Bottleneck in Post-training MI}
\begin{figure}[t]
\centering
\includegraphics[width=\linewidth]{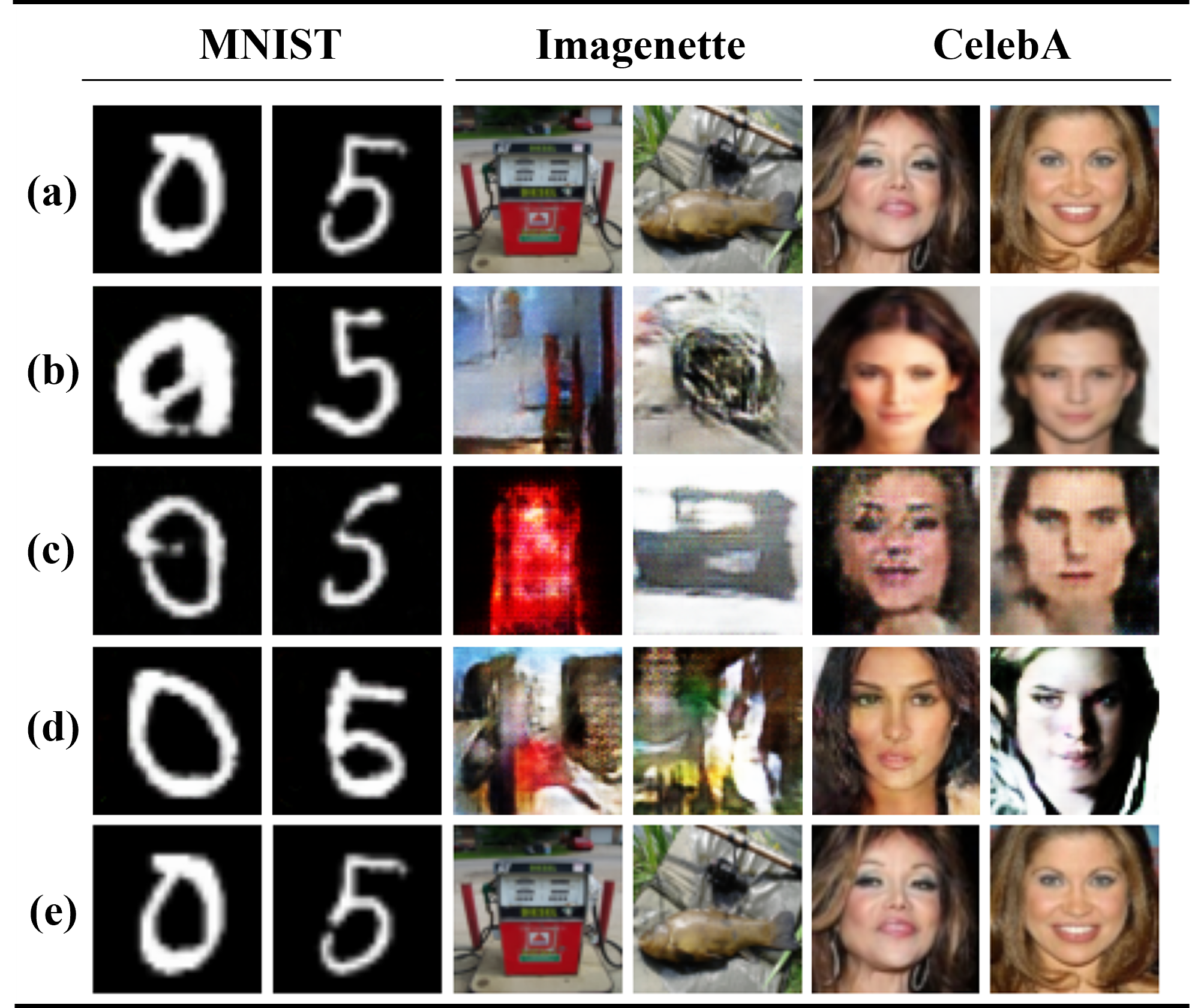}
\caption{Visual comparison of reconstructed samples across datasets. Rows: (a) ground truth, (b--d) post-training MI attacks (GMI, KEDMI, PPDG), and (e) GradLock (ours).}
\label{fig:main_results_mi}
\end{figure}

Table~\ref{tab:fidelity_mi_limits}, Table~\ref{tab:main_mi_limits_perf}, and Figure~\ref{fig:main_results_mi} present the quantitative comparisons and visual restoration results of the evaluated post-training MI attacks.
In this subsection, we focus on the MNIST dataset in the main text, with extended results for high-dimensional datasets (Imagenette and CelebA) provided in Appendix~\ref{sec:appendix_mi_results}.

\noindent \textbf{Quantitative Evaluation of Data Fidelity.}
As shown in Table~\ref{tab:fidelity_mi_limits}, post-training MI attacks exhibit an inverse correlation between dataset complexity and reconstruction fidelity.
On the \textbf{MNIST} dataset, these methods achieve an average SSIM of approximately 0.55, indicating that reconstructions capture coarse structures but lack fine-grained details. 
This limitation likely stems from their reliance on pre-trained feature representations, which are not optimized to preserve instance-specific information. 
As dataset complexity increases (e.g., Imagenette, CelebA), this gap becomes more pronounced; additional quantitative results are provided in Appendix~\ref{sec:appendix_mi_results}.

\noindent \textbf{Visual Reconstruction Analysis.}
As shown in Figure~\ref{fig:main_results_mi}, post-training MI attacks exhibit a strong dependency on the target model’s ability to learn robust feature representations.
On simple datasets like \textbf{MNIST}, where the model easily captures distinct digit shapes, these methods successfully recover recognizable inputs.
However, since these methods largely depend on the quality of the feature representations trained by the target model, they restore samples that are completely different from ground truth training data samples.
These representational bottlenecks are further exacerbated in high-dimensional datasets such as \textbf{Imagenette and CelebA}. 
Consequently, post-training MI attacks relying on this feature space struggle to resolve fine-grained details, resulting in smoothed or distorted artifacts that fail to capture the specific identity of the target. 
This demonstrates that the efficacy of post-training MI attacks is inherently bound by the generalization quality of the target model.

\noindent \textbf{Quantitative Comparison of Attack Performance.}
Table~\ref{tab:main_mi_limits_perf} highlights a key limitation of post-training MI attacks. 
While these methods achieve a high ASR on MNIST—reaching up to $0.95$ (KEDMI)—they simultaneously exhibit a large confidence gap, with values often exceeding $2{,}000$.
This substantial deviation from the original data distribution suggests that the reconstructed samples do not faithfully reflect the true feature distribution.
These results indicate that generative post-training MI attacks tend to optimize for classifier confidence using generic, model-aligned features, rather than recovering instance-specific characteristics of the training data. 

In conclusion, our results demonstrate that post-training MI attacks face significant hurdles in realistic deployment scenarios.

\subsection{Computational Efficiency}
Table~\ref{tab:inversion_time} highlights a critical disparity in computational overhead.
Post-training MI attacks like PPDG require prohibitive computation (e.g., 49 hours for a single batch on CelebA) due to iterative GAN fine-tuning.
In contrast, \textit{GradLock} achieves instant extraction (Avg. 0.377s) regardless of dataset size, which is significantly faster than the LSB encoding baseline (Avg. 2.957s).
This performance gap stems from the bit-level access and decompression overhead inherent in LSB encoding, whereas \textit{GradLock} directly extracts scaled raw pixel values.
These results confirm that our training-time approach enables scalable data exfiltration, overcoming the computational bottlenecks of traditional post-training MI attacks.

\input{tables/main_2-mi}

\input{tables/inversion_time}

%% file: tables/main_1-gradlock.tex
\begin{table*}[t]
\centering
\caption{Visual Fidelity Metrics of Training-time Injection Attacks across Datasets}
\label{tab:fidelity_injection}
\begin{tabular*}{\textwidth}{@{\extracolsep{\fill}} l c ccc ccc ccc }
\toprule
\multirow{2}{*}{Target Model} & \multirow{2}{*}{Method} & \multicolumn{3}{c}{MNIST} & \multicolumn{3}{c}{Imagenette} & \multicolumn{3}{c}{CelebA} \\
\cmidrule(lr){3-5} \cmidrule(lr){6-8} \cmidrule(lr){9-11}
& & SSIM$\uparrow$ & KNN$\downarrow$ & LPIPS$\downarrow$ & SSIM$\uparrow$ & KNN$\downarrow$ & LPIPS$\downarrow$ & SSIM$\uparrow$ & KNN$\downarrow$ & LPIPS$\downarrow$ \\
\midrule
\multirow{2}{*}{VGGNet-16} 
  & LSB Encoding & 1.000 & 0.000 & 0.000 & 1.000 & 0.000 & 0.000 & 1.000 & 0.000 & 0.000 \\
  & \textit{GradLock} (ours) & 1.000 & 0.009 & 0.000 & 1.000 & 0.011 & 0.000 & 1.000 & 0.011 & 0.000 \\
\midrule
\multirow{2}{*}{ResNet-18} 
  & LSB Encoding & 1.000 & 0.000 & 0.000 & 1.000 & 0.000 & 0.000 & 1.000 & 0.000 & 0.000 \\
  & \textit{GradLock} (ours)& 1.000 & 0.009 & 0.000 & 1.000 & 0.011 & 0.000 & 1.000 & 0.011 & 0.000 \\
\midrule
\multirow{2}{*}{DenseNet-121} 
  & LSB Encoding & 1.000 & 0.000 & 0.000 & 1.000 & 0.000 & 0.000 & 1.000 & 0.000 & 0.000 \\
  & \textit{GradLock} (ours) & 1.000 & 0.009 & 0.000 & 1.000 & 0.011 & 0.000 & 1.000 & 0.011 & 0.000 \\
\bottomrule
\end{tabular*}
\end{table*}

%% file: tables/main_2-gradlock.tex
\begin{table*}[t]
\centering
\caption{Performance Evaluation of Training-time Injection Attacks}
\label{tab:main_injection_perf}
\begin{tabular*}{\textwidth}{@{\extracolsep{\fill}} c c cc cc cc }
\toprule
\multirow{2}{*}{Target Model} &
  \multirow{2}{*}{Method} &
  \multicolumn{2}{c}{MNIST} &
  \multicolumn{2}{c}{Imagenette} &
  \multicolumn{2}{c}{CelebA} \\
  \cmidrule(lr){3-4} \cmidrule(lr){5-6} \cmidrule(lr){7-8}
  & & ASR $\uparrow$ & Conf. Gap $\downarrow$ & ASR $\uparrow$ & Conf. Gap $\downarrow$ & ASR $\uparrow$ & Conf. Gap $\downarrow$ \\
\midrule

\multirow{2}{*}{VGGNet-16}
  & LSB Encoding  & 1.00 & \textbf{0.000} & 0.82 & \textbf{0.000} & 0.44 & \textbf{0.000} \\
  & \textit{GradLock} (ours)   & 1.00 & \textbf{0.006} & 0.82 & \textbf{0.008} & 0.44 & \textbf{0.002} \\
\midrule

\multirow{2}{*}{ResNet-18}
  & LSB Encoding  & 1.00 & \textbf{0.000} & 0.65 & \textbf{0.000} & 0.49 & \textbf{0.000} \\
  & \textit{GradLock} (ours)   & 1.00 & \textbf{0.184} & 0.65 & \textbf{0.003} & 0.49 & \textbf{0.002} \\
\midrule

\multirow{2}{*}{DenseNet-121}
  & LSB Encoding  & 1.00 & \textbf{0.000} & 0.72 & \textbf{0.000} & 0.47 & \textbf{0.000} \\
  & \textit{GradLock} (ours)   & 1.00 & \textbf{0.042} & 0.72 & \textbf{0.004} & 0.47 & \textbf{0.011} \\
\bottomrule
\end{tabular*}
\end{table*}

%% file: tables/sub_generalizability.tex
\begin{table}[t]
\centering
\caption{Results of GradLock attack generalizability evaluation for Transformer-based architectures}
\label{tab:generalizability}
\begin{tabular*}{\linewidth}{@{\extracolsep{\fill}} c c ccc cc }
\toprule
\multirow{2}{*}{Dataset} & \multirow{2}{*}{Model} & \multicolumn{3}{c}{Visual Fidelity} & \multicolumn{2}{c}{Performance} \\
\cmidrule(lr){3-5} \cmidrule(lr){6-7}
& & SSIM$\uparrow$ & KNN$\downarrow$ & LPIPS$\downarrow$ & ASR$\uparrow$ & CG$\downarrow$ \\
\midrule
\multirow{2}{*}{MNIST} 
& \multirow{1}{*}{ViT} & 1.000 & 0.009 & 0.000 & 0.99 & 0.000 \\
& \multirow{1}{*}{SwinT} & 1.000 & 0.009 & 0.000 & 1.00 & 0.021 \\
\midrule
\multirow{2}{*}{Imagenette} 
& \multirow{1}{*}{ViT} & 1.000 & 0.011 & 0.000 & 0.60 & 0.000 \\
& \multirow{1}{*}{SwinT} & 1.000 & 0.011 & 0.000 & 0.83 & 0.001 \\
\midrule
\multirow{2}{*}{CelebA} 
& \multirow{1}{*}{ViT} & 1.000 & 0.011 & 0.000 & 0.44 & 0.001 \\
& \multirow{1}{*}{SwinT} & 1.000 & 0.011 & 0.000 & 0.73 & 0.003 \\
\bottomrule
\end{tabular*}
\end{table}

%% file: tables/statistical_analysis.tex
\begin{table}[t]
\centering
\caption{Average statistical features of weight distributions for Benign and GradLock-attacked (Ours) models on MNIST, Imagenette, and CelebA datasets.}
\label{tab:statistical_analysis}
\begin{tabular*}{\linewidth}{@{\extracolsep{\fill}} cccccc }
\toprule
Dataset & Method & Mean & Std.& Entropy & KS-norm \\
\midrule

\multirow{2}{*}{MNIST}
  & Benign & 0.033 & 0.221 & 4.855 & 0.031 \\
  & \textit{GradLock} &0.049 & 0.237 & 5.077 & 0.037 \\
\midrule

\multirow{2}{*}{Imagenette}
  & Benign  & 0.002 & 0.214 & 4.985 & 0.028 \\
  & \textit{GradLock}  & -0.003 & 0.106 & 4.539 & 0.031 \\
\midrule

\multirow{2}{*}{CelebA}
  & Benign  & -0.017 & 0.201 & 4.843 & 0.048 \\
  & \textit{GradLock}  & -0.019 & 0.221 & 4.968 & 0.036 \\
\bottomrule
\end{tabular*}
\end{table}

%% file: tables/main_1-mi.tex
\begin{table}[t]
\centering
\caption{Visual Fidelity Limits of Post-training Model Inversion (MI) Attacks}
\label{tab:fidelity_mi_limits}
\begin{tabular*}{\linewidth}{@{\extracolsep{\fill}} c c ccc }
\toprule
\multirow{2}{*}{Target Model} & \multirow{2}{*}{Method} & \multicolumn{3}{c}{MNIST} \\
\cmidrule(lr){3-5} 
& & SSIM$\uparrow$ & KNN$\downarrow$ & LPIPS$\downarrow$ \\
\midrule
\multirow{3}{*}{VGGNet-16} 
  & GMI   & 0.496 & 66.143 & 0.158 \\
  & KEDMI & 0.549 & 56.074 & 0.134 \\
  & PPDG  & 0.558 & 59.154 & 0.139 \\
\midrule
\multirow{3}{*}{ResNet-18} 
  & GMI   & 0.487 & 67.093 & 0.160 \\
  & KEDMI & 0.555 & 55.431 & 0.128 \\
  & PPDG  & 0.548 & 60.549 & 0.143 \\
\midrule
\multirow{3}{*}{DenseNet-121} 
  & GMI   & 0.501 & 65.971 & 0.162 \\
  & KEDMI & 0.554 & 55.494 & 0.133 \\
  & PPDG  & 0.544 & 60.379 & 0.140 \\
\bottomrule
\end{tabular*}
\end{table}

%% file: tables/main_2-mi.tex
\begin{table}[t]
\centering
\caption{Performance Limits of Post-training Model Inversion (MI) Attacks}
\label{tab:main_mi_limits_perf}
\begin{tabular*}{\linewidth}{@{\extracolsep{\fill}} c c cc }
\toprule
\multirow{2}{*}{Target Model} &
  \multirow{2}{*}{Method} &
  \multicolumn{2}{c}{MNIST} \\
  \cmidrule(lr){3-4}
  & & ASR $\uparrow$ & Conf. Gap $\downarrow$ \\
\midrule

\multirow{3}{*}{VGGNet-16}
  & GMI    & 0.74 & 151.025 \\
  & KEDMI  & 0.95 & 124.379 \\
  & PPDG    & 0.83 & 161.470 \\
\midrule

\multirow{3}{*}{ResNet-18}
  & GMI    & 0.77 & 2113.698 \\
  & KEDMI  & 0.94 & 1010.994 \\
  & PPDG    & 0.79 & 2047.647 \\
\midrule

\multirow{3}{*}{DenseNet-121}
  & GMI    & 0.73 & 564.894 \\
  & KEDMI  & 0.89 & 456.428 \\
  & PPDG   & 0.77 & 572.489 \\
\bottomrule
\end{tabular*}
\end{table}

%% file: tables/inversion_time.tex
\begin{table}[t]
\centering
\caption{Elapsed Time for Data Extraction (in seconds)}
\label{tab:inversion_time}
\begin{tabular}{@{}ccccc@{}}
\toprule
\multirow{2.5}{*}{Target Model} & \multirow{2.5}{*}{Method} & \multicolumn{3}{c}{Extraction Time (s)} \\
\cmidrule(lr){3-5}
 & & MNIST & Imagenette & CelebA \\
\midrule

\multirow{5}{*}{VGGNet-16}
  & GMI    & 824.65 & 812.18 & 916.46 \\
  & KEDMI  & 818.69 & 819.13 & 918.62 \\
  & PPDG  & 13004.61 & 8441.92 & 55434.96 \\
  & LSB Enc. & 3.86 & 4.56 & 3.63 \\
  & \textit{GradLock} & \textbf{0.61} & \textbf{0.61} & \textbf{0.67} \\
\midrule

\multirow{5}{*}{ResNet-18}
  & GMI    & 429.95 & 429.73 & 550.96 \\
  & KEDMI  & 423.68 & 423.67 & 552.65 \\
  & PPDG  & 11798.19 & 7001.58 & 48423.59 \\
  & LSB Enc. & 3.37 & 2.95 & 2.82 \\
  & \textit{GradLock} & \textbf{0.23} & \textbf{0.27} & \textbf{0.29} \\
\midrule

\multirow{5}{*}{DenseNet-121}
  & GMI    & 934.72 & 936.96 & 975.98 \\
  & KEDMI  & 743.45 & 746.35 & 946.98 \\
  & PPDG  & 23779.22 & 21527.76 & 177260.31 \\
  & LSB Enc. & 1.51 & 1.95 & 1.96 \\
  & \textit{GradLock} & \textbf{0.20} & \textbf{0.22} & \textbf{0.25} \\
\bottomrule
\end{tabular}
\end{table}

%% file: sections/V_conclusion.tex
\section{Discussion}

\noindent \textbf{The Threat of Silent Injection.}
We demonstrate that data leakage can be effectively induced through minimal intervention in the training pipeline, relying solely on simple weight manipulation within standard training frameworks.  
Remarkably, \textit{GradLock} preserves the original model performance across various datasets, thereby evading suspicion from end users or auditors.  
These results highlight a critical security risk in open-source environments, where training pipelines are widely reused, and pre-trained models are frequently redistributed without provenance tracking.

\noindent \textbf{Fidelity and Robustness Trade-off.}
Our results reveal the structural limitations of existing threats.
Conventional model inversion attacks struggle with representational bottlenecks, often failing to recover fine-grained details from capacity-constrained models.
Conversely, naive parameter encoding methods offer high fidelity in pristine states but lack robustness; they suffer catastrophic information loss under standard optimizations like quantization.
\noindent \textit{GradLock} bridges this gap, ensuring both high-fidelity extraction and persistence against real-world post-training optimization.

\noindent \textbf{Potential Countermeasures and Limitations.}
Despite its resilience, \textit{GradLock} is not without potential defenses. 
First, since our attack relies on deterministic indexing, invariance-based defenses such as weight permutation could disrupt the location mapping and hinder data recovery. 
%% (20260430) 여기 Second 내용은 앞의 실험과 반대되는 언급인데? 좀 건드려야겟네
%% (20260504) 살짝 톤만 다듬었습니다.
Second, explicit parameter manipulation may introduce subtle statistical deviations; therefore, highly rigorous weight distribution analysis or specialized modulation detection could potentially flag the presence of injected data.
%Second, explicit parameter manipulation creates statistical anomalies; therefore, rigorous weight distribution analysis or modulation detection could potentially flag the presence of injected data.

%% (20260504) 문장 순서도 앞뒤 바꿔서 좀더 자연스럽게 배치했습니다.
However, as demonstrated in Section 4.6, \textit{GradLock} effectively operates within the natural variance of stochastic optimization across most training environments, rendering such statistical audits practically challenging under real-world constraints.
Furthermore, constructing automated detectors capable of identifying these subtle patterns typically requires a large and diverse set of both clean and attacked reference models, which is rarely available in practical supply-chain scenarios. 

\noindent \textbf{Future Directions.}
Future work should explore stealthier, spread-spectrum embedding strategies to obscure these statistical footprints and mitigate overhead.
Ultimately, our findings emphasize that securing the AI supply chain requires moving beyond simple performance checks to implementing comprehensive auditing tools capable of detecting deep parameter-level anomalies.

\section{Conclusion}
We present \textit{GradLock}, a novel training-time injection attack that embeds sensitive training samples directly into the weights of deep neural networks.
By leveraging deterministic indexing and selective gradient masking, \textit{GradLock} enables covert data injection that is both functionally and visually stealthy.  
Extensive experiments demonstrate that \textit{GradLock} exhibits superior robustness and persistence compared to existing training-time injection baselines. 
Most importantly, it maintains effective data recovery even under rigorous model compression scenarios—such as quantization and pruning—where naive parameter encoding methods suffer catastrophic information loss.
Our user-centered deployment study further provides preliminary qualitative evidence of the real-world feasibility of such attacks, revealing that most participants in our study failed to detect or question malicious training pipelines.
These findings underscore an urgent need for more robust auditing tools, provenance tracking, and runtime inspection to secure modern machine learning workflows.
As the AI ecosystem increasingly relies on pre-trained models and third-party training code, attacks like \textit{GradLock} highlight a new frontier of model-based data extraction that demands both technical and procedural countermeasures.

%% file: sections/appendix.tex
\appendix

\section{GradLock Training Procedure} \label{sec:appendix_algorithm}

We formally describe the complete training and extraction procedure of \textit{GradLock} in Algorithm~\ref{alg:gradlock}. 
The process is divided into three phases: (1) Deterministic Indexing \& Injection, (2) Training with Gradient Locking, and (3) Post-training Data Extraction.

\begin{figure*}[t]
    \centering
    \includegraphics[width=0.7\linewidth]{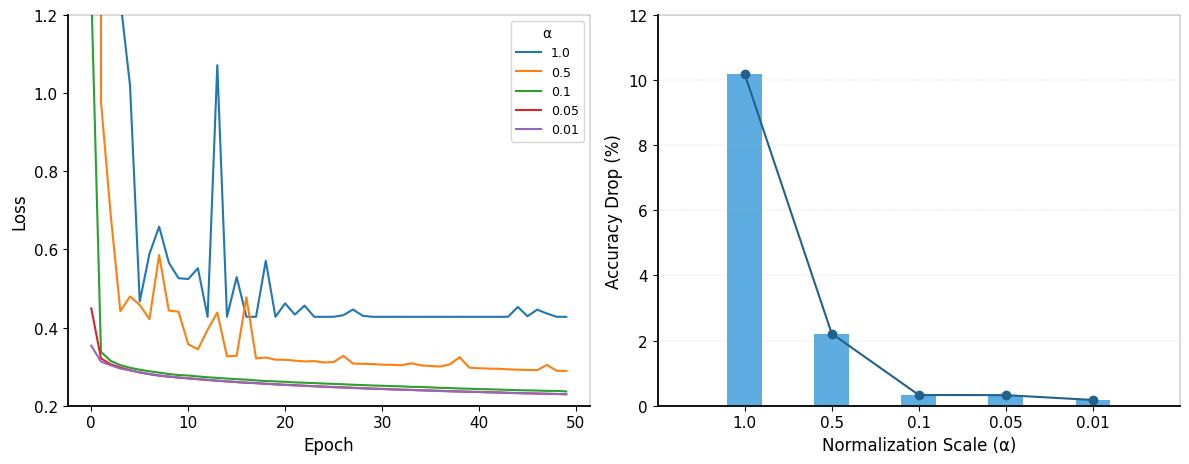}
    \caption{Effect of normalization scale on training loss (left) and test accuracy drop rate (right).}
    \label{fig:norm_scale_analysis}
\end{figure*}

\begin{algorithm}[h]
\caption{GradLock: Training-time Injection and Post-training Extraction}
\label{alg:gradlock}
\begin{algorithmic}[1]
\REQUIRE Target model $f_\theta$, Private dataset $\mathcal{D}_{priv}$, Target layer weights $W \in \mathbb{R}^{N_{row} \times d}$, Injection ratio $\rho$, Scaling factor $\alpha$
\ENSURE Extracted data $\mathcal{D}_{ext}$

\STATE \textbf{Phase 1: Deterministic Indexing \& Injection}
\STATE $N_{row} \leftarrow \text{Rows}(W)$
\STATE $\mathcal{R} \leftarrow \{ \lfloor N_{row} \cdot |\sin(\frac{\pi i}{N_{row}/\rho})| \rfloor \mid i \in [0, \lfloor \rho N_{row} \rfloor - 1] \}$ \COMMENT{Eq. 1: Select row indices}
\STATE $M \leftarrow \mathbf{0}_{size(W)}$; $M[\mathcal{R}, :] \leftarrow 1$ \COMMENT{Binary Mask for Locking}
\FOR{$i \in [0, |\mathcal{R}|-1]$}
    \STATE $x_{priv} \leftarrow \mathcal{D}_{priv}[i]$
    \STATE $v_{flat} \leftarrow \texttt{Flatten}(x_{priv})$
    \STATE $W[\mathcal{R}_i, :] \leftarrow \alpha \cdot \text{Normalize}(v_{flat})$ \COMMENT{Eq. 4: Inject Payload}
\ENDFOR

\STATE \textbf{Phase 2: Training with Gradient Locking}
\FOR{epoch $= 1$ to $E$}
    \FOR{batch $(x, y)$ in $\mathcal{D}_{train}$}
        \STATE $\hat{y} \leftarrow f_\theta(x)$ \COMMENT{Eq. 3: Forward Pass (Active use of payload)}
        \STATE $\mathcal{L} \leftarrow \text{LossFunction}(\hat{y}, y)$
        \STATE $g \leftarrow \nabla_W \mathcal{L}$ \COMMENT{Compute Gradients}
        \STATE $g \leftarrow g \odot (1 - M)$ \COMMENT{Eq. 2: Mask Gradients (Locking)}
        \STATE $W \leftarrow \text{OptimizerStep}(W, g)$ \COMMENT{Update Remaining Parameters}
    \ENDFOR
\ENDFOR

\STATE \textbf{Phase 3: Data Extraction (Post-training)}
\STATE $\mathcal{D}_{ext} \leftarrow \emptyset$
\FOR{$i \in [0, |\mathcal{R}|-1]$}
    \STATE $w_{rec} \leftarrow W[\mathcal{R}_i, :]$
    \STATE $\hat{x} \leftarrow \texttt{Reshape}(w_{rec} / \alpha)$ \COMMENT{Eq. 5: Recover Data}
    \STATE $\mathcal{D}_{ext} \leftarrow \mathcal{D}_{ext} \cup \{\hat{x}\}$
\ENDFOR
\RETURN $\mathcal{D}_{ext}$
\end{algorithmic}
\end{algorithm}

\section{Details of Ablation Studies} \label{sec:appendix_ablation}
In this appendix, we provide a detailed analysis of \textit{GradLock}, specifically focusing on the normalization scaling factor ($\alpha$) and the numerical breakdown of the gradient locking ratio ($\rho$).

\input{tables/ablation_1}

\subsection{Impact of Normalization Scale on Optimization Stability}
As discussed in Section~\ref{experiment}, the scaling factor $\alpha$ plays a pivotal role in stabilizing the training process. 
Direct injection of normalized data (in the range $[-1, 1]$) into weight parameters—which typically follow a distribution with a much smaller standard deviation—can introduce significant noise gradients during backpropagation.

\noindent \textbf{Training Loss Convergence.}
Figure~\ref{fig:norm_scale_analysis} (Left) illustrates the training loss trends across different $\alpha$ values.
\begin{itemize}
    \item \textbf{High $\alpha$ ($\alpha \in \{1.0, 0.5\}$):} The loss curve exhibits high variance and instability. The magnitude of the injected values dominates the gradient updates, preventing the optimizer from settling into a smooth minimum.
    \item \textbf{Low $\alpha$ ($\alpha \le 0.1$):} The loss curves (green, red, purple lines) converge smoothly, exhibiting behavior almost identical to the baseline training process. This indicates that the scaled injection acts as a negligible perturbation to the optimization trajectory.
\end{itemize}

\input{tables/main_1-mi2}

\noindent \textbf{Accuracy Degradation.}
Figure~\ref{fig:norm_scale_analysis} (Right) quantifies the impact on test accuracy.
With $\alpha=1.0$, the model suffers a catastrophic accuracy drop ($>10\%$) due to the distortion of the weight distribution.
However, as $\alpha$ decreases, the accuracy recovery is substantial.
The degradation stabilizes at a minimum around $\alpha=0.05$ and remains consistently negligible through $\alpha=0.01$.
Based on these empirical results, we selected $\alpha=0.01$ as the optimal operating point, ensuring that the injection remains stealthy to performance monitoring while retaining sufficient precision for data recovery.

\subsection{Detailed Sensitivity Analysis of Gradient Locking}
Table~\ref{tab:rho_deact_acc} provides the precise numerical breakdown of the trade-off between the locking ratio ($\rho$) and model performance.

The \textbf{Locked Parameters (\%)} column demonstrates the non-linear relationship between $\rho$ and the actual frozen parameters. 
Due to the properties of the sine-based deterministic indexing function, index collisions occur more frequently as $\rho$ increases.
This natural collision mechanism prevents the adversary from linearly monopolizing the parameter space, inadvertently acting as a regularizer that preserves model capacity even at high injection attempts.
Notably, at our chosen default of $\rho=0.5$, less than half of the layer's weights are locked, securing space for 146 images (as shown in the Capacity column) while resulting in a negligible accuracy drop of 0.20\%.

\section{Detailed Protocol for User Deployment Study} \label{sec:appendix_user_study}

To ensure the validity and reproducibility of the practical deployment analysis presented in Section~\ref{experiment}, we provide the detailed protocol used in our user study.

\noindent \textbf{Participants.}
We targeted 30 undergraduate seniors majoring in Software Engineering. All participants had completed a basic course in Artificial Intelligence and were proficient in Python, but they were not security experts. 
This demographic reflects potential developers active in collaborative open-source environments.

\noindent \textbf{Task Description.}
Participants were enrolled in an "Information Security Application" course and assigned a term project to develop a digital forensic tool. Specifically, they were instructed to train a CNN-based binary classifier to identify a specific target person (e.g., a celebrity) from a large image dataset. They were provided with a baseline training repository (`detect-target-cnn`) containing the necessary scripts and were graded primarily on the model's classification performance (F1-score, Accuracy).

\input{tables/main_2-mi2}

\noindent \textbf{The Malicious Payload.}
The provided codebase included a file named `utils.py`, which ostensibly contained helper functions for model setup. 
The \textit{GradLock} logic was concealed within a function named `initialize()`, masquerading as a standard initialization routine (e.g., Xavier or Kaiming initialization). This function, called automatically at the start of training, executed the gradient masking and data injection logic under the guise of setting up initial parameters.

\noindent \textbf{Evaluation Criteria.}
We monitored three key behaviors to assess security awareness:
\begin{enumerate}
    \item \textbf{Inspection:} Did the user open and read `utils.py` before execution?
    \item \textbf{Detection:} Did the user identify the gradient masking logic within the initialization function as anomalous?
    \item \textbf{Submission:} Did the user submit the trained model file (`.pth`) generated by the compromised script?
\end{enumerate}

The high submission rate (93.3\%) confirms that in goal-oriented development environments, utility preservation (i.e., the model training successfully) often overrides security vigilance, especially when malicious code is camouflaged as standard initialization procedures.

\section{Extended Results on Post-training MI Attacks}
\label{sec:appendix_mi_results}
Table~\ref{tab:fidelity_mi_limits2} and Table~\ref{tab:main_mi_limits2} provide a detailed quantitative breakdown of post-training MI attacks across diverse architectures. 
Unlike the moderate success observed on the MNIST dataset, these high-dimensional datasets (Imagenette and CelebA) exhibit a catastrophic collapse in reconstruction fidelity, with SSIM values frequently dropping below 0.10 and LPIPS values exceeding 0.40. 
Furthermore, the excessively large confidence gaps—often surpassing 2,000—underscore a profound distributional divergence, confirming that the reconstructed samples drift significantly from the true feature manifold of the training data.
These extended results further validate the information bottleneck hypothesis discussed in Section 4. While generative baselines may achieve a high ASR by synthesizing model-aligned features from external priors, they fail to recover instance-specific identities, instead converging toward blurry, class-average templates. 
This performance gap is particularly evident in the high KNN distances, which demonstrate that post-training MI attacks primarily satisfy the classifier's decision boundaries without extracting the granular, private samples that \textit{GradLock} successfully preserves.

\section{Open Science}

The source code, trained models, and experimental scripts for reproducing the results of GradLock are publicly available at:

\url{https://github.com/KimJinSeong-Git/GradLock.git}

The repository includes:
\begin{itemize}
    \item Full implementation of the GradLock attack
    \item Training and evaluation scripts
    \item Configuration files and pretrained models used in the experiments
    \item Scripts to reproduce the main tables and figures presented in the paper
\end{itemize}

For the post-training model inversion baselines (GMI, KEDMI, and PPDG), we adapted the official implementations by extracting and integrating their core inversion logic. The adapted baseline code is available at:

\url{https://github.com/KimJinSeong-Git/GradLock.git}

Due to repository storage limits, large-scale public datasets (Imagenette and CelebA) are not hosted directly in the repository.

%% file: tables/ablation_1.tex
\begin{table}[t]
\centering
\caption{Effect of $\rho$ on gradient locking, storage capacity, and accuracy degradation (\%).}
\label{tab:rho_deact_acc}
\begin{tabular}{@{}c c c c c@{}}
\toprule
$\rho$ & \makecell{Locked\\Gradient (\%)} & \makecell{\textbf{Capacity}\\\textbf{(Images)}} & \makecell{Test \\ Accuracy (\%)} & \makecell{Dropped \\ Accuracy (\%)} \\
\midrule
0.1 & 9.69 & 28 & 88.750 & -0.138 \\
0.2 & 19.49 & 58 & 88.771 & -0.114 \\
0.3 & 29.29 & 87 & 88.737 & -0.152 \\
0.4 & 39.08 & 116 & 88.672 & -0.226 \\
\textbf{0.5} & \textbf{48.98} & \textbf{146} & \textbf{88.695} & \textbf{-0.201} \\
0.6 & 58.52 & 175 & 88.555 & -0.358 \\
0.7 & 66.94 & 200 & 88.285 & -0.661 \\
0.8 & 74.08 & 221 & 88.207 & -0.749 \\
0.9 & 79.49 & 237 & 88.234 & -0.719 \\
1.0 & 82.50 & 246 & 88.097 & -0.873 \\
\bottomrule
\end{tabular}
\end{table}

%% file: tables/main_1-mi2.tex
\begin{table*}[t]
\centering
\caption{Visual Fidelity Limits of Post-training Model Inversion (MI) Attacks}
\label{tab:fidelity_mi_limits2}
\begin{tabular*}{\textwidth}{@{\extracolsep{\fill}} l c ccc ccc ccc }
\toprule
\multirow{2}{*}{Target Model} & \multirow{2}{*}{Method} & \multicolumn{3}{c}{MNIST} & \multicolumn{3}{c}{Imagenette} & \multicolumn{3}{c}{CelebA} \\
\cmidrule(lr){3-5} \cmidrule(lr){6-8} \cmidrule(lr){9-11}
& & SSIM$\uparrow$ & KNN$\downarrow$ & LPIPS$\downarrow$ & SSIM$\uparrow$ & KNN$\downarrow$ & LPIPS$\downarrow$ & SSIM$\uparrow$ & KNN$\downarrow$ & LPIPS$\downarrow$ \\
\midrule
\multirow{3}{*}{VGGNet-16} 
  & GMI   & 0.496 & 66.143 & 0.158 & 0.094 & 84.059 & 0.411 & 0.053 & 81.569 & 0.284 \\
  & KEDMI & 0.549 & 56.074 & 0.134 & 0.100 & 80.963 & 0.422 & 0.214 & 72.655 & 0.228 \\
  & PPDG  & 0.558 & 59.154 & 0.139 & 0.083 & 81.272 & 0.444 & 0.173 & 78.463 & 0.229 \\
\midrule
\multirow{3}{*}{ResNet-18} 
  & GMI   & 0.487 & 67.093 & 0.160 & 0.089 & 83.402 & 0.421 & 0.053 & 82.339 & 0.285 \\
  & KEDMI & 0.555 & 55.431 & 0.128 & 0.092 & 81.972 & 0.447 & 0.200 & 71.665 & 0.261 \\
  & PPDG  & 0.548 & 60.549 & 0.143 & 0.062 & 82.571 & 0.409 & 0.183 & 79.588 & 0.229 \\
\midrule
\multirow{3}{*}{DenseNet-121} 
  & GMI   & 0.501 & 65.971 & 0.162 & 0.097 & 84.106 & 0.416 & 0.057 & 79.776 & 0.279 \\
  & KEDMI & 0.554 & 55.494 & 0.133 & 0.082 & 84.453 & 0.444 & 0.210 & 73.030 & 0.245 \\
  & PPDG  & 0.544 & 60.379 & 0.140 & 0.061 & 80.561 & 0.401 & 0.188 & 80.945 & 0.219 \\
\bottomrule
\end{tabular*}
\end{table*}

%% file: tables/main_2-mi2.tex
\begin{table*}[t]
\centering
\caption{Performance Limits of Post-training Model Inversion (MI) Attacks}
\label{tab:main_mi_limits2}
\begin{tabular*}{\textwidth}{@{\extracolsep{\fill}} c c cc cc cc }
\toprule
\multirow{2.5}{*}{Target Model} &
  \multirow{2.5}{*}{Method} &
  \multicolumn{2}{c}{MNIST} &
  \multicolumn{2}{c}{Imagenette} &
  \multicolumn{2}{c}{CelebA} \\
  \cmidrule(lr){3-4} \cmidrule(lr){5-6} \cmidrule(lr){7-8}
  & & ASR $\uparrow$ & Conf. Gap $\downarrow$ & ASR $\uparrow$ & Conf. Gap $\downarrow$ & ASR $\uparrow$ & Conf. Gap $\downarrow$ \\
\midrule

\multirow{3}{*}{VGGNet-16}
  & GMI    & 0.74 & 151.025 & 0.96 & 59.318 & 0.01 & 8.586 \\
  & KEDMI  & 0.95 & 124.379 & 0.68 & 58.041 & 0.07 & 7.604 \\
  & PPDG    & 0.83 & 161.470 & 0.89 & 63.173 & 0.52 & 10.055 \\
\midrule

\multirow{3}{*}{ResNet-18}
  & GMI    & 0.77 & 2113.698 & 0.85 & 21.708 & 0.00 & 10.087 \\
  & KEDMI  & 0.94 & 1010.994 & 0.76 & 24.490 & 0.08 & 8.442 \\
  & PPDG    & 0.79 & 2047.647 & 0.85 & 23.903 & 0.71 & 10.163 \\
\midrule

\multirow{3}{*}{DenseNet-121}
  & GMI    & 0.73 & 564.894 & 0.79 & 22.204 & 0.04 & 37.911 \\
  & KEDMI  & 0.89 & 456.428 & 0.78 & 24.009 & 0.09 & 43.944 \\
  & PPDG   & 0.77 & 572.489 & 0.81 & 23.258 & 0.83 & 41.472 \\
\bottomrule
\end{tabular*}
\end{table*}